\documentclass[usenatbib,useAMS,usegraphicx]{mn2e}

\usepackage{times} %% For PDF
\newcommand\arcs{\mbox{$^{\prime\prime}\!$}}
\newcommand\arcm{\mbox{$^\prime$}}
\def\lsim{~\raise0.3ex\hbox{$<$}\kern-0.75em{\lower0.65ex\hbox{$\sim$}}~}
\def\gsim{~\raise0.3ex\hbox{$>$}\kern-0.75em{\lower0.65ex\hbox{$\sim$}}~}
\newcommand{\rab}{$R_{\rm AB}$}
\newcommand{\hkab}{$HK^\prime_{\rm AB}$}
\newcommand{\mum}{$\,\mu$m}

\newcommand{\chandra}{\textsl{Chandra}}
\newcommand{\hst}{\textsl{HST}}
\newcommand{\sirtf}{\textsl{Spitzer}}

\def\mnras{MNRAS}
\def\apj{ApJ}
\def\aap{A\&A}
\def\aj{AJ}

\begin{document}
\title[The HDF-N SCUBA Super-map II]
{The HDF-North SCUBA Super-map II: Multi-wavelength properties}

\author[Borys et. al.]{
\parbox[t]{\textwidth}{
\vspace{-1.0cm}
Colin Borys$^{1}$,
Douglas Scott$^{2}$,
Scott Chapman$^{1},$
Mark Halpern$^{2}$,
Kirpal Nandra$^{3}$,
Alexandra Pope$^{2}$
}
\vspace*{6pt}\\
$^{1}$ California Institute of Technology, Pasadena, CA 91125, USA\\
$^{2}$ Department of Physics \& Astronomy, University of British Columbia,
       Vancouver, BC, V6T 1Z1, Canada \\
$^{3}$ Astrophysics Group, Imperial College London, Blackett Laboratory,
Prince Consort Road, London SW7 2AZ, UK\\
\vspace*{-0.5cm}}

\date{Submitted 27 April 2004}

\maketitle

\begin{abstract}
We present radio, optical and X-ray detected counterparts to the sub-mm
sources found using SCUBA in the Hubble Deep Field North region (GOODS-N).
A new counterpart identification statistic is developed to identify 
properties of
galaxies detected at other wavelengths that can be used to aid counterpart
identification.  We discriminate between criteria that can use used to
pre-select sub-mm bright objects, and those that identify the counterpart to a
known sub-mm object.  Optically faint galaxies detected in the deepest
$1.4\,$GHz radio continuum maps are the only effective way of pre-selecting
SCUBA galaxies, and
radio sources are the best way to identify counterparts to known sub-mm
detections.   Looking at radio spectral indices, only the steeper
sources (indicative of star formation) are detected in the sub-mm.  Although
we find several X-ray identifications, we show that deep \chandra\ images do 
not contribute to counterpart
identifications, since in all cases they are already detected in the more
easily obtained VLA radio maps.  We also find find no evidence for
clustering between \chandra\ and SCUBA sources in this field. 
For a known SCUBA position, the reddest source tends to be the correct
association, although we can find no cut on colour, magnitude, or clustering
property that efficiently pre-selects for SCUBA sources. 
$15\,\mu$m {\sl ISO\/} sources are statistically detected by SCUBA, but the
limiting mid-IR flux is not low enough to provide useful constraints.
We present postage stamp strips for each SCUBA detection in separate
bands from X-ray to radio, providing direct visual evidence that
approximately half of the sub-mm sources in this field remain
unidentified, despite an abundance of deep multi-wavelength data.
\end{abstract}

\begin{keywords}
methods: statistical -- methods: numerical -- galaxies: formation --
infrared: galaxies -- galaxies: starburst
\vspace*{-1.25cm}
\end{keywords}

\section{Introduction}
Finding distant IR luminous galaxies is now a routine occurrence for the
Sub-millimetre Common User Bolometer Array \citep[SCUBA][]{1999MNRAS.303..659H}
 on the
James Clerk Maxwell Telescope (see \citealt{2002PhR...369..111B}). 
Specifically, we estimate that there have been about 300 published 
detections in the first 5 years since the instrument was commissioned in 1997.
Much more challenging is determining \textsl{what} the galaxies are.  To get 
a complete picture it is necessary to
compare the sub-mm sources against data obtained at other wavelengths, but this
is often quite difficult for a number of reasons.  First, the very dust
responsible for the strong IR emission obscures and re-processes light at other
wavelengths, making the objects more difficult to detect.  Secondly, the 
negative K-correction in the sub-mm that allows an almost distance-independent
ability to detect these galaxies does not apply in the optical and radio. The
most serious issue, however, is the large
beam-size in sub-mm surveys; often there are several sources within the SCUBA
beam detected at other wavelengths, and without some means to discriminate
among them, it is uncertain which (if any) of the sources is the host of the
the far-IR emission.

Until the advent of the next generation of sub-mm interferometers (particularly
ALMA), there is not much that can be done to significantly improve the
resolution of the images.  But we can address the other issues by obtaining
very deep images of SCUBA fields at other wavelengths in order to better
identify faint counterparts. The HDF-N 
\citep{1996AJ....112.1335W,2000ARA&A..38..667F} and its flanking fields, with 
its wide range of deep imaging across many wavebands, is 
arguably the best region to use for this purpose.  

In this paper, we use this rich multi-wavelength data-set to determine
counterparts to the SCUBA detected sources in a roughly
$0.05\,{\rm deg}^2$ region centred on the HDF-N, as described in 
\citet[][hereafter called `Paper I']{2003MNRAS.344..385B}
Given the volume of data still being
obtained in the region due to the the Great Observatories Origins Deep Survey
(GOODS) programme, we choose to include only data published or obtained prior
to the deep ACS images released in August 2003.   A larger sub-mm map of the
HDF-N, and the study of other GOODS data is reserved for future work.

\section{Review of the HDF-N sub-mm maps}
Sub-mm maps of the HDF-N were obtained using the SCUBA camera
at the James Clerk Maxwell Telescope, including more than 60 shifts of
telescope time.  We combined the various data-sets into a single map, which
we refer to as the Super-map.  Combining this with detailed simulations allowed us
to extract and assess the reliability of the sources.
The full data reduction and source extraction algorithm are described in
Paper~I.

In summary, the Super-map contains 19 sources which were detected at 850\mum\ 
with a signal-to-noise ratio (SNR) $>4$.  An additional list of 15 sources 
was presented between 3.5 and 4$\sigma$.  Monte-Carlo simulations of the 
data suggest that the 4$\sigma$ catalogue should have at most one false 
detection, but that the number of spurious sources rises drastically for 
lower thresholds.  Thus the $3.5-4\sigma$ catalogue is considered less 
reliable, although we later argue that at least a third are real.  The reality 
of the 5 sources detected at 450\mum\ is much less secure however, since none
were found coincident with a source at 850\mum.

Subsets of the data used here have been published by other groups, and
in Paper~I we compared them against our map, finding no serious discrepancies.  However, 
a new survey by \citet{rant} appeared after this work was submitted which takes a more
critical view of other SCUBA analyses of the HDF-N. In Appendix B we compare our results
with theirs, and find that in fact their results support the work here and in previous
publications.

\section{Statistical criteria for finding counterparts}
\label{sec:statid}
To identify the most likely counterparts for our SCUBA detections, we need
first to choose a method for associating counterparts.  Since the Super-map has
non-uniform noise, our catalogue includes sources at a variety of signal and
SNR levels (including some faint high SNR sources, as well as brighter low
SNR sources).  Because of the variable effects of pointing uncertainty, noise
and source confusion, we choose to adopt a single approach for {\it all\/}
our sources, namely to use a fixed search radius.  Next we have to choose a
value for this radius, outside of which we reject sources.  We are confident 
that our Super-map {\em as a whole} has astrometry reliable to within
3\arcs, since pointing corrections were rarely much larger, and in addition
because the stacked radio/sub-mm flux density begins to drop if the map is
shifted by more than this (see section \ref{sec:radio}).  Since some parts of 
the map are dominated by single observations, we need to consider the typical 
2\arcs\ pointing uncertainty of the JCMT as well. 

The Monte-Carlo simulations described in Paper~I indicate an additional
uncertainty of at most 5\arcs\ for the simulated sources recovered at 850\mum,
which is caused by a combination of confusion and residual sky noise altering
the centroid of these faint sources. Adding these in quadrature leads to a
conservative search radius of 7\arcs, which is comparable to values chosen by
other groups.  Although this choice of search radius is somewhat arbitrary, 
we now show that it is in some sense an `optimal' value.

\subsection{An alternative to the $\bmath{P}$--statistic}
Given the large positional uncertainty for SCUBA sources, several objects from
an image in another waveband with higher resolution are possible counterparts 
to a given SCUBA source.  But which one is the correct ID?  A measure often adopted by the 
sub-mm community is the so-called `$P$--statistic'
(e.g.~\citealt{1986MNRAS.218...31D}).  A different approach, and one we 
advocate and derive next, is to compare the  {\em ensemble} of SCUBA 
positions to see which types of counterparts are most likely associated with 
the sub-mm detections.

Given a surface density of some class of object of $n$ per unit area, the 
random probability that one or more lies within a distance $\theta$ of a 
specific SCUBA source is
\begin{equation}
\label{equ:probchance}
P=1-\exp(-\pi n \theta^2).
\end{equation}
This is the traditional $P$--statistic, and the lower its value,
the less likely it is that the object is associated with
a source by chance.  Table~\ref{tab:pstat} lists the radii within which there
is a 95 per cent chance that the object is the counterpart to the SCUBA galaxy.
It can be misleading however.
For example, if an ERO was found 9.3\arcs\ away from a SCUBA source, the
$P$--statistic would suggest there is only a 5 per cent chance it is not
the correct counterpart.  However, this object should have been rejected 
outright because it lies well outside the estimated 7\arcs\ uncertainty in 
the SCUBA position.

Identifying counterparts between catalogues of objects at different
wavelengths is an endeavour with a long history in the radio and
X-ray communities (see for example
\citealt{1977A&AS...28..211D,1983MNRAS.204..355P,1991ApJS...76..813S}).
The `$P$--statistic' is
not the only approach used, and there are many discussions in the literature
of how to use
special properties of candidate counterparts, how to include astrometric
uncertainties, cuts on flux ratios, etc.  However, these discussions tend to
be focused entirely on identifying each individual source.

Since the detailed nature of SCUBA-bright galaxies is still largely unknown,
then when comparing with other wavelength images,
it is also useful to have a statistic which assesses the identification
of the SCUBA sample as a whole.  Such a statistic is also easier
to interpret, particularly if we choose a fixed search radius.  For what
follows we assume a uniform probability of association over a 7\arcs\ radius
circle and zero outside.  We can then
straightforwardly estimate the probability of finding a set of objects within
the search radius using Poisson statistics, as follows.

Given a population with a surface density of $n$, the probability of finding
{\em no} sources within a distance, $\theta$, of a given point is
$p_0(\theta)=\exp(-\pi n \theta^2)$ (for a derivation see
e.g.~\citealt{1989MNRAS.241..109S}).  This is just the same
statement as in equation~(\ref{equ:probchance}), and deals with the statistics
of a single object.  The probability that no counterparts are found for $M$
independent searches (e.g.~in our case for $M$ separate SCUBA sources) is
\begin{equation}
p_{0,M}(\theta)=\Pi_1^Mp_0(\theta) = p_0^M(\theta).
\end{equation}
Similarly, the probability of finding a single counterpart is
\begin{equation}
p_{1,M}(\theta)=M\times p_0(\theta)^{M-1} {\left(1-p_0(\theta)\right)}.
\end{equation}
The pre-factor $M$ is needed because there are $M$ different ways to pick a 
single object from a set of $M$ objects. Using a similar argument, the 
probability that of the $M$ points, $K$ have at least one counterpart is 
\begin{equation}
p_{K,M}(\theta)={{M!}\over{(M-K)!K!}}p_0(\theta)^{M-K}
 {\left(1-p_0(\theta)\right)}^K.
\end{equation}
Therefore, of $M$ objects, the probability that $K$ or more of them have at
least one counterpart is
\begin{equation}
p_{K+}=\sum_{i=K}^M p_{i},
\end{equation}
where we have dropped the $M$ and $\theta$.  Note that $p_0 + p_{1+} \equiv 1$.

We have applied this statistic to our data and present the results in 
Table~\ref{tab:pstat}.  Note how striking some of the implications are.  For 
instance, the probability that 11 of the 19 SCUBA sources have a 1.4 GHz 
radio source within 7\arcs\ just by chance is essentially zero. 

Still, one has to treat this statistic with some degree of caution. For 
example, these probabilities assume that the populations are unclustered; as 
we will comment on later in this paper, clustering evidence does exist for 
some of the populations we compare the SCUBA sources against. We simply state 
this caveat for now, and will describe clustering results on a case-by-case 
basis later.

Finally, is the choice of a 7\arcs\ search radius a good one?  For larger radii,
one will get more counterparts, but the probability of more matches by chance
will also increase. Therefore a trade-off exists between search radius and probability,
and by calculating $p_{K+}$ as a function of radius, we indeed find a minimum
at 7\arcs.  This is independent of which catalogue is used to compute the 
statistic.

\begin{table*}
\caption{Statistical measures for assessing the reliability of counterparts.  
For each of the catalogues we use in this paper, we tabulate several
quantities. The first, $\theta_{5\%}$ is the separation out to which
there is a 5 per cent probability that a source is not the correct 
counterpart.  
The second, $p_{0}$ should be read as `the random probability that none of 
the $M$ SCUBA sources have at least one counterpart within our adopted 
7\arcs\ search radius'. We then give the number $K$ of $M$ SCUBA sources that 
have an identifications within 7 \arcs. Note that 
some of the catalogues do not extend over the entire region of the Super-map.
Finally we give $p_{K+}$, the probability of $K$ or more of the $M$ SCUBA
objects having at least one counterpart.  We present these statistics both for
our $>4\sigma$ SCUBA sources and for the full $>3.5\sigma$ catalogue.
Here, LBG means `Lyman Break Galaxy' and ERO is an `Extremely Red Object'.
}
\label{tab:pstat}
\begin{tabular}{lllllllll} 
\hline
Class                           &  Number within &$\theta_{5\%}$ & \multicolumn{3}{c}{$>4\sigma$ sources}   & \multicolumn{3}{c}{$>3.5\sigma$ sources}   \\
                                &  survey area   &  arcsec       & $p_{0}$ & $K/M$  & $p_{K+}$            & $p_{0}$ & $K/M$   & $p_{K+}$            \\\hline\hline
VLA 1.4\,GHz                    &  135           &  8.8          &  0.54   & 11/19 & $p_{10+} < 10^{-10}$  & 0.33    & 14/34 & $p_{14+} < 10^{-12}$ \\
VLA 8.5\,GHz                    &   51           & 14.2          &  0.79   & 6/19 & $p_{6+} < 10^{-7}$  & 0.66    &  7/34 & $p_{7+}  < 10^{-6}$\\
ISO 15\mum                      &   99           &  4.1          &  0.26   & 4/9  & $p_{4+} = 0.03$     & 0.14    &  6/13  & $p_{6+}  < 10^{-2}$\\
LBG                             &  132           &  5.6          &  0.38   & 1/12 & $p_{1+} = 0.62$     & 0.23    &  1/18 & $p_{1+}  = 0.77$    \\
\chandra\ 2\,Ms                 &  328           &  5.4          &  0.20   & 8/19 & $p_{8+} < 10^{-4}$  & 0.06    &  14/34 & $p_{14+}  < 10^{-6}$ \\
\rab$<24$                       & 2626           &  2.0          &  0.00   & 9/19 & $p_{9+} = 0.59$     & 0.00    &  17/34 & $p_{17+}  = 0.45$    \\   
\rab$<22$                       &  454           &  4.8          &  0.12   & 1/19 & $p_{1+} = 0.88$     & 0.02    &  3/34 & $p_{3+}  = 0.70$    \\   
\rab$\,{-}\,$\hkab$>3.00$       &  801           &  3.7          &  0.03   & 4/19 & $p_{4+} = 0.41$     & 0.00    &  7/34 & $p_{7+}  = 0.36$    \\
\rab$\,{-}\,$\hkab$>3.93$ (ERO) &  121           &  9.3          &  0.58   & 0/19 & $p_{0+} = 1.00$     & 0.37    &  1/34 & $p_{1+}  = 0.63$    \\\hline
\end{tabular}
\end{table*}

\subsection{Statistical measures of sub-mm flux density from known objects}
One can compare a catalogue of objects against a sub-mm image via a `stacking'
analysis in order to get a sense of the average
sub-mm properties of the sample.  The procedure is to take a list of detected objects from a survey,
make cuts of some sort (if desired), and then sum up the flux density from a
map at the positions of all the objects. Specifically, we take 

\begin{equation}
{\bar{S}={{\sum_i{S_i}\sigma_i^{-2}}\over{\sum_i\sigma_i^{-2}}}},
\left(d\bar{S}\right)^2=\sum_i\sigma_i^{-2},
\end{equation}
where $S_i$ and $\sigma_i$ are the sub-mm flux density and error estimates at 
the position of object $i$.  This technique is not restricted to sub-mm
maps; for instance, \citet{2002ApJ...576..625N} have compared a sample of LBGs
against \chandra\ X-Ray maps.  

%One can estimate the contribution of a given 
%class of object to the 850\mum\ background by multiplying the stacked
%average by the number density of the sources.  This can be compared against 
%the measured value of $8.6$--$11.4\,{\rm mJy}\,{\rm arcmin}^{-2}$
%\citep{1996A&A...308L...5P,1998ApJ...508...25H,1998ApJ...508..123F}.

It is important to check for systematic effects in these analyses.  We take
the list of 51 stars in the HDF-N region from the work of
\cite{1998A&A...333..106M} and correlate them against the HDF Super-map. One
would expect no signal from these stars as they are not sub-mm emitters. The
stacked average, $-0.20\pm0.15\,$mJy, is consistent with both the average value
of the map (0.02\,mJy), and the distribution of fluxes derived from many
realizations of taking 51 random positions in the map.   Therefore we can
proceed with some assurance that any significant stacked signal is real.

Another effect to consider is the variation in the number of objects
being compared.  For large $N$, the stacked flux density will approach the
average value of the map if the objects are distributed randomly on the sky.  
Therefore the stacked flux density from $N_1$ objects
cannot be directly compared to that from $N_2$ objects if they are significantly
different (unless a correction is made).  Hence in our stacking analyses we only make comparisons among sub-samples with
equal number of objects.

\section{Multi-wavelength database and sub-mm pre-selection criteria}
\label{sec:catdescriptions}
We have obtained source catalogues and images of the HDF-N 
over a wide range of wavelengths that are relevant to
understanding the nature of the sources detected in the sub-mm maps.  Before
discussing the sub-mm objects individually, we present an overview of each
catalogue and explain their relation to the sub-mm population.  In
particular, we will search for any criterion that make it possible to pre-select
sub-mm detections in our map.  We will start
with the radio associations, which, as seen in Table~\ref{tab:pstat}, are
certainly the most important link to the SCUBA population.

\subsection{VLA and WSRT radio observations}
\label{sec:radio}
The entire region of our Super-map has been imaged to roughly
$9\,\mu\rm{Jy}$ RMS using both
the VLA \citep{2000ApJ...533..611R} and the WSRT \citep{2000A&A...361L..41G}
radio telescopes at 1.4\,GHz .   A smaller area survey, covering
$11.3\arcm\times11.3\arcm$ surrounding the central HDF region, was conducted
with the VLA at 8.5\,GHz \citep{1998AJ....116.1039R} to an RMS of
$1.6\,\mu\rm{Jy}$.  We have obtained the catalogues from each survey and the
VLA maps at both frequencies in order to perform comparisons with our sub-mm
map.  Within the HDF-N Super-map there are 135 1.4\,GHz VLA sources detected
at $>5\sigma$.  In the combined 8.5\,GHz catalogues (the primary list plus some
fainter, less securely detected objects), there are 51 sources, 26 of which
have a 1.4\,GHz counterpart within 3\arcs.  We note that the WSRT and VLA
1.4\,GHz catalogues generally agree, except that the WSRT positions are, on
average, 1.5\arcs\ West of the VLA positions.  Unless otherwise noted, we 
will use the VLA catalogue.

Within the 7\arcs\ search radius, 11(6) of the 19 objects from the Super-map
$4\sigma$ list have a 1.4(8.5)\,GHz radio counterpart.  7 of these radio
sources are detected at both 1.4 and 8.5\,GHz.  As shown in
Table~\ref{tab:pstat}, the chance of 11 SCUBA objects having a nearby
radio source just at random is less than $10^{-10}$. Based on the this result
we will assume that a radio source within 7\arcs\ is the correct SCUBA counterpart,
unless other compelling evidence excludes it.

\subsubsection{Radio source stacking analysis}
\label{sec:radio_stacking}
When we stack all the 1.4 GHz radio positions on the Super-map, we calculate an
average 850\mum\ flux density of $1.8\pm0.1\,$mJy.  This is a very significant
detection, and comparable to results from \cite{2000AJ....119.2092B} and
\cite{2002ApJ...570..557C}.  To see if there are correlations between radio
objects and the sources we \textsl{do not} detect, we mask out a circular
region of radius 8\arcs\ (the half-width of the beam size at 850\mum) from each of the
detected sub-mm sources in the full $>3.5\sigma$ catalogue.  The $850\,\mu$m
flux density stacked
at the radio positions is still $S_{850}=0.6\pm0.1\,$mJy, suggesting
that although the bulk of the radio-stacked flux density
comes from the sub-mm detected sources, additional flux density associated
with radio galaxies which are weakly
detected in the sub-mm is still present.
Stacking the 8.5\,GHz sources yields an average 850\mum\ flux density of
$S_{850}=1.7\pm0.1\,$mJy, and drops to
$0.3\pm0.2$ when the SCUBA sources are masked out.

\begin{figure}
\begin{center}
\includegraphics[width=3.0in,angle=0]{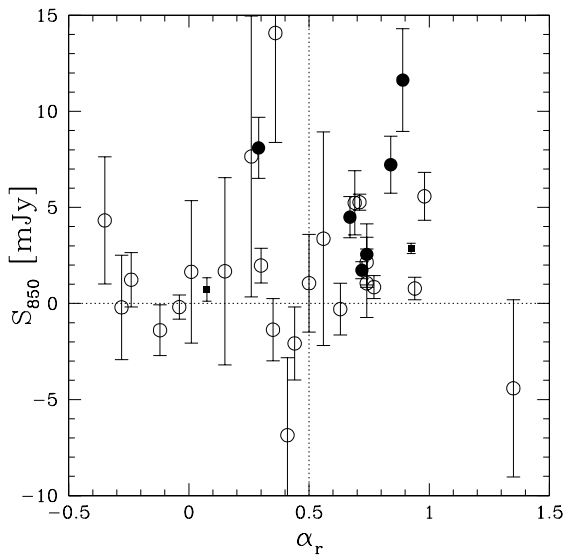}
\caption{Comparing SCUBA 850\mum\ flux density against the radio spectral
index, $\alpha_{\rm r}$.  Open circles are sources undetected with SCUBA,
while solid ones represent radio galaxies found within 7\arcs\ of a sub-mm
source from the $>4\sigma$ catalogue.  Solid squares represent the stacked
flux density in 2 bins of $\alpha_{\rm r}$ (with each bin containing 15 sources). 
}
\label{fig:radioindex}
\end{center}
\end{figure}

Radio galaxies are often characterised by the slope of their spectrum
($f_\nu\propto\nu^{-\alpha_{\rm r}}$). Roughly speaking, an inverted spectrum
($\alpha_{\rm r}<0$) indicates the presence of self-absorbed synchrotron
radiation from an AGN.  Flat spectra ($0<\alpha_{\rm r}<0.5$) also suggest
self-absorption and AGN, but in addition can be due to increased high frequency
radio emission from star-formation.  Steeper values of the index are associated
with diffuse synchrotron radiation from star-forming galaxies
\citep{1992ARA&A..30..575C}, with the `canonical' index being 0.8.  In
Fig.~\ref{fig:radioindex} we plot the 850\mum\ flux density
at the position of the 30
galaxies detected at both 1.4 and 8.5\,GHz (see Table~5 in Richards 2000)
against the radio spectral index. The sample with $\alpha_{\rm r}<0.5$ is not
significantly detected in the sub-mm, but the `star-forming' sources are.

\subsubsection{Registering sub-mm images using radio data}
\begin{figure}
\begin{center}
\includegraphics[width=3.0in,angle=0]{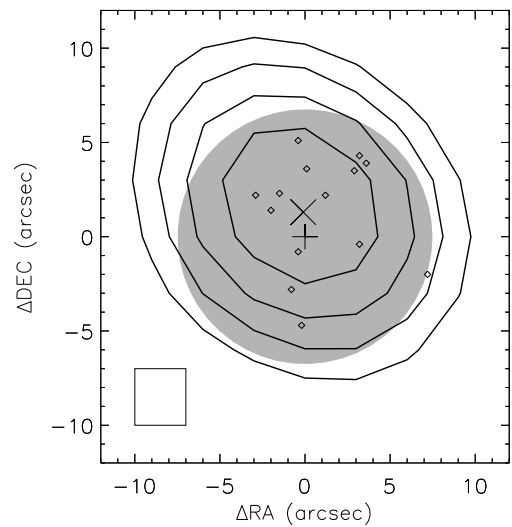}
\caption{Using the VLA radio sources to test the sub-mm astrometry.
We shift the radio map relative to the Super-map and calculate the stacked 
850\mum\ flux density.  Contours are drawn at 0.5, 0.6, 0.7, and 0.8 times
the peak value of the stacked flux density.  The grey circle is the same
size as the SCUBA beam (FWHM), while the black square is the size of a pixel
in the sub-mm Super-map
(3\arcs\,).  Note that the contour at 0.5 is essentially the FWHM of the
correlation distribution, and thus can be directly compared to the grey circle.
The contours are more extended due to a combination of astrometry shifts in the
Super-map and clustering.  The
contours prefer a centre (shown by the cross symbol) that is offset by roughly
1.5\arcs\ (half a pixel) from the unshifted sub-mm map (denoted by the plus
symbol).  Diamonds denote the offsets for the 14 radio sources which we later
claim are the correct identifications to sub-mm sources in the HDF-N.
}
\label{fig:radiopoint}
\end{center}
\end{figure}

Based on the strength of the sub-mm/radio correlation we can test the astrometry
of the sub-mm map by shifting it and re-calculating the stacked flux density
at the
radio positions.  In Fig.~\ref{fig:radiopoint} we plot the result.  The
correlation (as determined by the stacked flux) decreases quickly for distances
greater than a SCUBA beam-size.  A slight offset of 1.5\arcs\ is seen between 
the
VLA and SCUBA maps with a consistent shift found using only the positions of 
the radio counterparts to our SCUBA sources. Although no result in this 
paper is sensitive to such a small offset, we make the shift in order to 
have the best astrometry for the SCUBA Super-map. As a consequence, some
of the `SMMJ' source names have changed by 1 digit from the list in Paper 1.
Note that the correlation in Fig.~\ref{fig:radiopoint} seems to extend past 
the size of a theoretical,
diffraction limited SCUBA beam.  This would be the case if individual scans had
small offsets between them and were then co-added.  Fitting the FWHM of the
brightest sources in the Super-map does show that some objects have
profiles up to 15 per cent wider than the nominal 15\arcs\ beam.  However, it
is also possible that there is a contribution from clustering of the radio
sources.  As noted in \citet{2000ApJ...533..611R}, there is a $>5\sigma$
detection of radio source clustering on scales of $\sim 0.1$--$2.0\arcm$.

\subsection{\chandra\ X-Ray imaging}
ACIS, the Advanced CCD Imaging Spectrometer \citep{1987ApL&C..26...35N} on
board \chandra\ performed a 2\,Msec integration on a $17\arcm\times17\arcm$
region
surrounding the HDF-N, making it the deepest X-ray observation
yet obtained.  The survey reaches $0.5-2.0\,$keV (soft) and $2-8\,$keV (hard)
flux limits of about $3\times10^{-20}$ and $2\times10^{-19}$ W\,m$^{-2}$,
respectively \citep{2003AJ....126..539A}.
To examine the relationship between SCUBA sources and X-ray flux,
we use the \chandra\ HDF-N catalogue of \cite{2003AJ....126..539A} that covers
our entire sub-mm map.  The catalogue lists 503 objects, 451 in the soft band
and 332 in the hard band, with many sources detected in both.  The positions
are accurate to within 0.3\arcs\ near the centre of the field, but the
uncertainty increases to 1.5\arcs\ near the edge.  Of these detections, 328
fall within our sub-mm map, and 8 of the 19 SCUBA sources have an X-ray source
within our 7\arcs\ search radius (3 of these have 2 X-ray sources within 
7\arcs\,). 

\subsubsection{\chandra\ stacking analysis}
\begin{figure}
\begin{center}
\includegraphics[width=3in,angle=0]{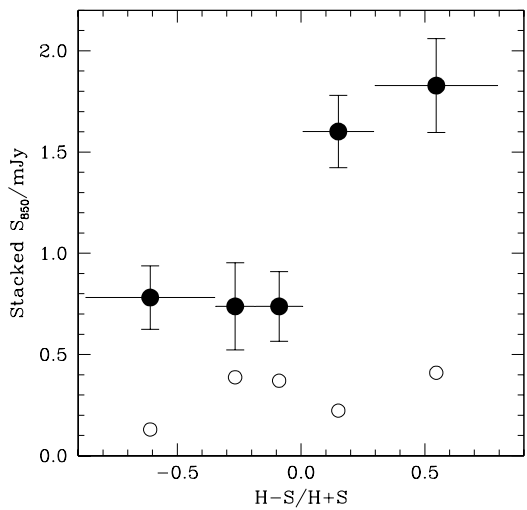}
\caption{Stacked sub-mm flux density as a function of X-Ray hardness ratio.
We have
split the \chandra\  sources into 5 equal sized bins, with roughly 65 objects
per bin.  Solid circles show binned averages, with horizontal bars denoting
the size of the bins.  The open circles (with error bars omitted for clarity)
are the same but with the detected SCUBA sources masked out. 
}
\label{fig:stackchandra}
\end{center}
\end{figure}

We can characterise the spectral shape of the X-ray emission via the hardness
ratio, $(H-S)/(H+S)$.  For this purpose it is conventional to use the counts
in the hard ($H$) and soft ($S$) bands instead of the fluxes.
Fig.~\ref{fig:stackchandra} reveals a strong correlation between those sources
with a hard X-ray spectrum and stacked sub-mm flux density.  We obtain an overall average 850\mum\ flux density
of $S_{850}=1.0\pm0.1\,$mJy. Most of this signal comes from the
hardest third of the sources, although even the softer ones are significantly
detected. In general the stacked flux density is due to SCUBA detected objects,
and not from X-ray detected objects that are faint in the sub-mm.  This is in
stark contrast to the radio stacking results.

Within our 7\arcs\ search radius we find 8 \chandra\ sources out of 19
${>}\,4\sigma$ SCUBA sources, and 14 in the full ${>}\,3.5\sigma$ catalogue.
However, with over twice as many X-ray detected objects (in the \chandra\
catalogue) than radio objects, the probability of a chance
occurrence is much higher.  None the less, the probability of this many matches
is $<10^{-4}$ (as shown in Table~\ref{tab:pstat}), and therefore highly
significant.  This estimate could be biased if the SCUBA sources
cluster around X-ray galaxies, as we now discuss.

\subsubsection{The clustering of sub-mm and X-ray sources}
\citet[][2003]{almainiconf} claim to detect a strong clustering signal between
SCUBA and X-ray sources in the UK 8--mJy survey \citep{2002MNRAS.331..817S}.
They determined that although only 1 of 17 SCUBA sources in the 8--mJy
catalogue had a genuine \chandra\ detected counterpart, X-ray and SCUBA sources
tend to be found close together.   The clustering signal at small angular
separations can be due to objects which are the same galaxy identified at both
wavelengths.  However, a positive correlation out to $\sim1\arcm$ must be due
to the two populations tracing out the same large scale structure.  Hence the
8--mJy results imply a spatial link between X-ray bright and sub-mm bright
populations.  It has been speculated that these two catalogues trace the same
population, but at different stages in their evolution.  This would explain why
there is no enhanced overlap between SCUBA and X-ray sources, yet a clustering
signal between the two populations can still be detected.  High redshift
clusters cover angles $\sim1\arcm$, which is consistent with this
picture. 

\begin{figure}
\begin{center}
\includegraphics[width=3.0in,angle=0]{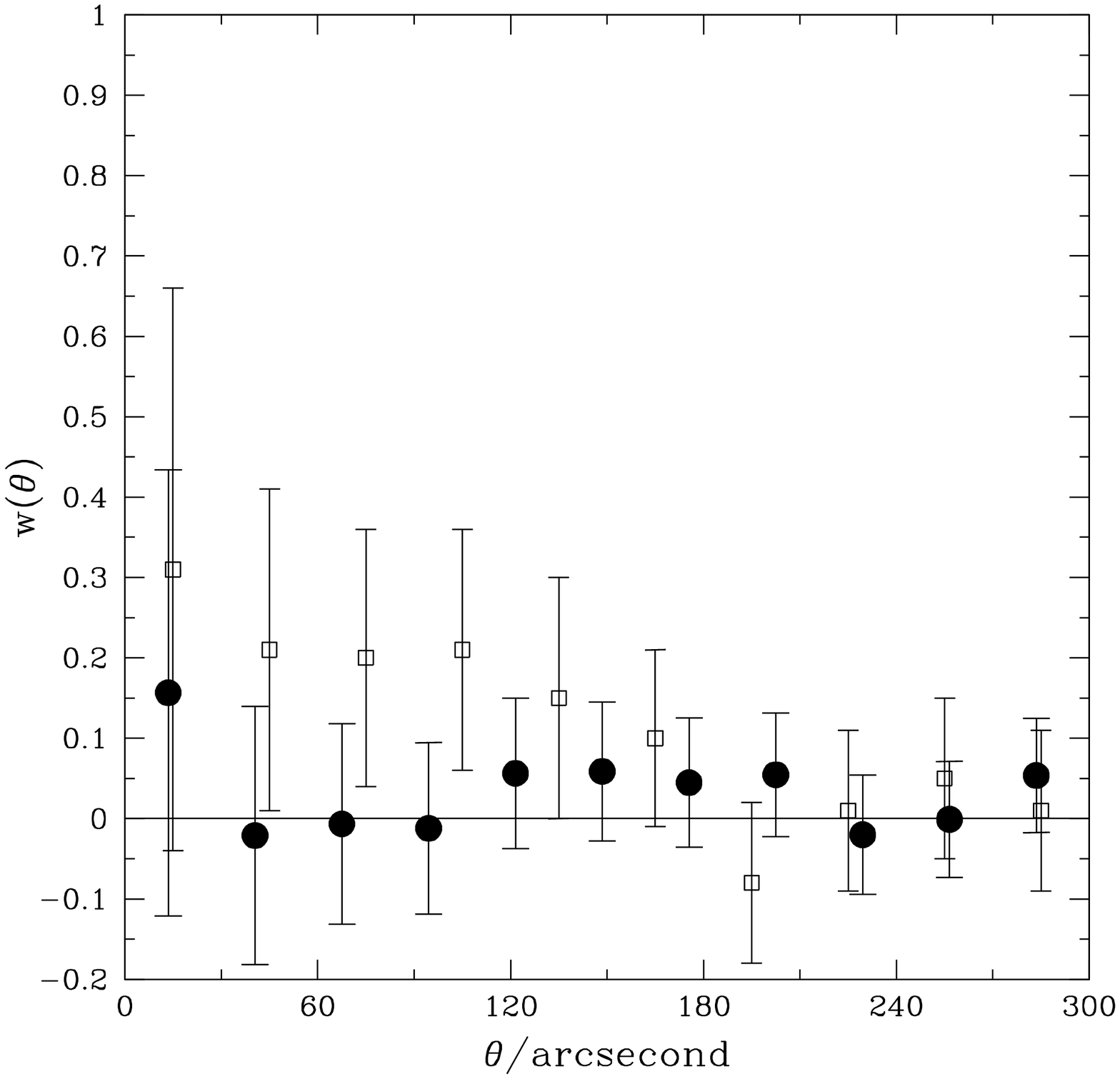}
\caption[Clustering between SCUBA and X-ray detected objects.]{Clustering
between SCUBA and X-ray detected objects.  Solid circles show the
cross-clustering estimate based on our $>4\sigma$ sources and the published
X-ray coordinates from \chandra. No detection is seen. Open squares are
{}from \citet{almainiconf}, which show some weak evidence of clustering, though
using only the \chandra\ 1\,Ms data and the shallower HDF-N SCUBA map in
\citet{2002MNRAS.330L..63B}. We have ignored X-ray sources closer than
7\arcs, since those are generally the SCUBA sources themselves.
}
\label{fig:scuxray}
\end{center}
\end{figure}

We performed a clustering analysis for sources found within the HDF-N
Super-map using the statistic \citep{1993ApJ...417...19H}
\begin{equation}
\label{equ:wsx}
w(\theta) = {{SX\times R_X R_S}\over{SR_X\times XR_S}} - 1.
\end{equation}
Here, $w(\theta)$ is the angular correlation function and the pairs of 
sources are counted between different catalogues:
$S$ and $X$ represent SCUBA and X-ray sources, while $R_X$ and $R_S$ are random
X-ray and SCUBA catalogues.
 Monte-Carlos are required to generate the random
catalogues.  In these simulations, we assume that the sensitivity to X-ray
sources is the same across the entire field, and can therefore place sources
randomly on the sky using simple Poisson statistics.  This is not precisely
true, since the X-ray sensitivity decreases with off-axis distance.  However,
the sources we are using are quite significantly detected and therefore we can
safely assume that we are not biasing our results.  We cannot make the same
assumption for the SCUBA detected galaxies, since the sensitivity is far from
uniform across the field.  Therefore we generated 500 simulated maps using the
sub-mm source count model described in Paper I.   Using these random catalogues
and the estimator in equation \ref{equ:wsx}, we calculated $w(\theta)$ for
30\arcs\ wide angular bins, and plot the results in Fig.~\ref{fig:scuxray}.  No
clustering signal is found.

Why is there a discrepancy between the 8--mJy results and those here?  First
we note that of the 7 SCUBA sources with an X-ray source nearby, only 3 are
brighter than the sensitivity limit of the X-ray observations in the 8--mJy
region.  Thus we repeated the analysis ignoring X-ray sources fainter than the
8--mJy sensitivity, but still no correlation was found.  One idea is that the
measured clustering signal is due to gravitational lensing of SCUBA sources
behind the foreground large scale structures that host the X-ray objects
\citep{2003MNRAS.338..303A}.  If this were true, one might expect a stronger
signal when using only the brightest (and presumably more strongly lensed)
SCUBA sources, since the steepness of the source-counts here leads to a
stronger bias in the number of objects detected.   We therefore re-did the
clustering analysis using only the $>7\,$mJy sources but still found no
detection.  Even including the $>3.5\sigma$ detections to increase the sample
size did not help.

We conclude there is no evidence of an X-ray/SCUBA clustering correlation in
the HDF, and speculate that this may be due to field-to-field variation.
On the other hand, \citet{almainiconf} only report weak clustering in their HDF-N
analysis. We also note that \citet{2003MNRAS.338..303A} used a
different $w(\theta)$ estimator than the one given in equation~(\ref{equ:wsx}).
This estimator is not symmetric between the sub-mm and X-ray, and
thus may be prone to bias.

In defense of the foreground lensing hypothesis, there is one SCUBA object,
SMMJ123621+621252, which has 7 \chandra\ detections within 20\arcs\, none of
which are the counterpart to the SCUBA source.  This object is discussed in
the next section.  There are also 2 SCUBA sources in a crowded field (based on
optical images) that have an apparent \chandra\ ID with a spectroscopic
redshift, though the redshift is too low to be consistent with the radio-FIR 
(Far Infrared)
correlation.  These examples may indicate systems where foreground sources
lens the background SCUBA galaxy.  However, these are anecdotal
suggestions, and we stress that we could find no statistical evidence
for such associations.

\subsubsection{Are \chandra\ sources a good way to identify SCUBA-bright
galaxies?}
\label{sec:chandragood}
Of the 8 SCUBA galaxies that have a \chandra\ detected galaxy within 7\arcs,
only two were not first detected in the 1.4\,GHz radio map: SMMJ123637+621155
and SMMJ123656+621201.  In both cases, there is a radio detection that we call
the counterpart, but the X-ray point source is not coincident with it. A third
object, near SMMJ123652+621225 does have a radio ID, but (as explained in the
next section) is not the correct counterpart.

Thus we conclude that all secure SCUBA IDs that have a \chandra\ detection
are also
seen in the 1.4\,GHz radio maps.  Indeed, the radio is much more efficient,
with a success rate here of 10/18 compared with 5/18 objects (ignoring
the peculiar object SMMJ123652+621225).  The count 
rates for these 5
\chandra\ sources are such that only 1 would have been detected by \chandra\
in an exposure time equivalent to that used for the VLA 1.4\,GHz image
($\sim$50 hours).  So, per unit integration time, it is currently much more
efficient to find SCUBA counterparts with the VLA than with \chandra.

\subsubsection{X-ray properties of SCUBA objects}
Including sources from the supplementary SCUBA catalogue, we
present a list of 10 X-ray detected SCUBA objects in Table~\ref{tab:xrayprops};
five from the $>4\sigma$ list and five from the supplemental catalogue.
Only 4 are in common with the list of 7 presented in
\citet{2003AJ....126..539A}.  From that work, which used the previous
releases of HDF-N sub-mm data, we have rejected SMMJ123622+621618 
because its large distance from the sub-mm galaxy renders it an unlikely ID.
SMMJ123618+621552 is in a region of extended X-ray flux, and we were unable to 
determine if a point source was present. \citet{2003AJ....126..539A} re-reduce
the data using different detection parameters and do find a source here, but
we choose to employ only the original $2\,$Ms catalogue.
Neither of the two \chandra\ galaxies 
in the vicinity of  SMMJ123713+621204 stand out as the correct ID, 
so we reject those as well. 

For these \chandra\ identifications we calculated radio and X-ray luminosities
using the formulae in \citet{2003AJ....125..383A}.  We assumed a radio spectral
index of 0.8 and an unobscured photon index of $\Gamma=2$ for each source.
These results are summarized in Table \ref{tab:xrayprops} and 
Fig.~5.

The luminosities of most of the sources
lie within the range determined by \citet{2002AJ....124.2351B} and 
\citet{2001ApJS..137..139S} for local star-forming galaxies.  Four
sources lie above the relation however, suggesting
an AGN component. \citet{2003AJ....125..383A} use templates of
various galaxies and quasars to argue that, although these systems harbour
an AGN, their FIR luminosity is still dominated by 
star-formation \citep[see also][]{2003MNRAS.343..585F}.
Better constraints on the FIR luminosity (via observations at other sub-mm
wavelengths) along with optical spectroscopy could be used to strengthen this
argument, but such data are not yet available for a significant sample.

\begin{table*}
\caption{X-ray properties of SCUBA sources with secure identifications.
All X-ray quantities are obtained from the \chandra\ 
2\,Ms catalogue \citep{2003AJ....126..539A}.  Fluxes are in units of
${\rm 10^{-18} W\,m^{-2}}$. `FB', `SB' and `HB' refer to the full, soft and
hard bands, respectively.  Redshifts are taken from 
Table \ref{tab:hdfredshift}, and luminosities are calculated using
the prescription in \citet{2003AJ....125..383A}.
}
\label{tab:xrayprops}
\begin{tabular}{lllllllllll}
\hline
SMM ID         & X-Ray                  & Hardness & Effective      & \multicolumn{3}{c}{X-Ray flux}                                           & $S_{1.4}$ &$z$ & $\log[L_{\rm X}]$     & $\log[L_{1.4}]$ \\
%\cline{2-4}\cline{8-10}
               & Counts                 & Ratio    & $\Gamma$       & \multicolumn{1}{c}{FB} & \multicolumn{1}{c}{SB} & \multicolumn{1}{c}{HB} & $(\mu$Jy) &    & (erg\,s$^{-1}$) & (W\,Hz$^{-1}$)      \\\hline\hline
\multicolumn{11}{c}{850\mum\ detections $>4\sigma$}\\\hline
J123616+621516 & $130.4$    & $ -0.09$ & $1.0^{+0.2}_{-0.2}$ & 1.02 & $ 0.18$ & $ 0.80$ & $53.9\pm8.4$   & $2.06$              & $24.1$        & $43.5$       \\
J123645+621449 & $12.6$     & $< 0.00$ & $1.4$               & 0.08 & $ 0.03$ & $ 0.14$ & $124.0\pm9.8$  & $1.9^{+1.0}_{-0.7}$ & $24.4\pm0.4$  & $42.3\pm0.5$ \\
J123650+621316 & $27.2$     & $-0.39$  & $1.4$               & 1.64 & $ 0.61$ & $<1.23$ & $49.2\pm7.9$   & $0.475$             & $22.6$        & $42.1$       \\       
J123701+621146 & $14.1$     & $<-0.18$ & $1.4$               & 0.09 & $ 0.04$ & $<0.12$ & $128.0\pm9.9$  & $1.52$              & $24.2$        & $42.1$       \\
J123707+621410 & $84.4$     & $ +0.24$ & $0.4^{+0.2}_{-0.2}$ & 0.98 & $ 0.09$ & $ 0.91$ & $45.3\pm7.9$   & $3.7^{+2.8}_{-1.5}$ & $24.6\pm0.4$  & $44.1\pm0.6$ \\\hline
\multicolumn{11}{c}{Additional 850\mum\ detections $>3.5\sigma$}\\\hline
J123607+621019 & $59.1$     & $>+0.36$ & $0.1^{+0.1}_{-0.1}$ & 0.74 & $<0.05$ & $ 0.77$ & $74.4\pm9.0$   & $0.47$              & $22.8$        & $41.8$       \\
J123608+621431 & $37.6$     & $<-0.22$ & $1.4$               & 0.25 & $ 0.08$ & $<0.22$ & $68.9\pm8.8$   & $0.472$             & $22.7$        & $41.3$       \\
J123628+621046 & $198.4$    & $ +0.34$ & $0.2^{+0.2}_{-0.2}$ & 2.31 & $ 0.17$ & $ 2.23$ & $81.4\pm8.7$   & $1.013$             & $23.6$        & $43.1$       \\
J123635+621237 & $53.3$     & $<-0.42$ & $1.6^{+1.6}_{-1.6}$ & 0.29 & $ 0.10$ & $<0.17$ & $230\pm14$     & $1.219$             & $24.2$        & $42.4$       \\
J123652+621352 & $20.4$     & $<-0.03$ & $1.4$               & 0.13 & $ 0.03$ & $<0.14$ & $<45$          & $1.355$             & $<23.6$       & $42.2$       \\\hline
\end{tabular}
\end{table*}

\begin{figure}
\label{fig:rxlumi}
\includegraphics[width=3in,angle=0]{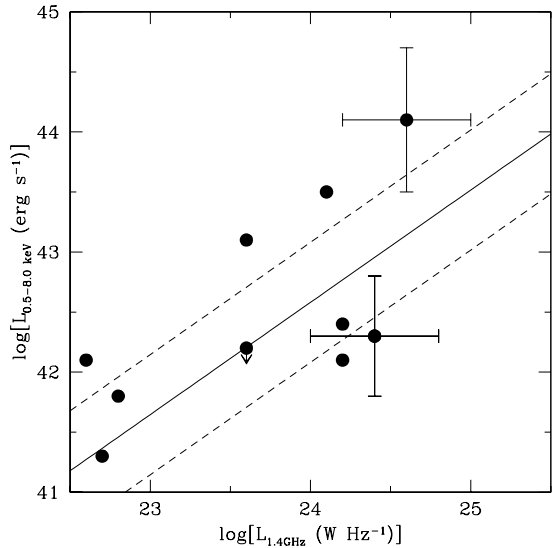}
\caption[Radio/X-ray luminosity relation for secure SCUBA detections.]{
Radio/X-ray luminosity relation for secure SCUBA detections.  The solid 
and dashed lines represent the best fit (and $\pm1\sigma$ range) of the 
relation as derived by \citet{2002AJ....124.2351B} for a sample
of local galaxies \citep{2001ApJS..137..139S}.
}
\end{figure}

\subsection{Optical-NIR imaging}
The most comprehensive published optical survey, aside from the \hst\ imaging itself,
is presented in \citet{2004AJ....127..180C}.  They obtained deep optical and
near-IR (NIR) images covering the entire HDF-N Super-map in the
$U,B,V,R,I,z^\prime$ and
$HK^\prime$ bands, and made the images
and catalogues publicly accessible via their web page.  We use their $R$ and
$z^\prime$ selected catalogues (consisting of almost 49,000 objects) which
cover a
0.2 deg$^2$ region.  The sub-mm map is completely contained within this area.
The catalogue uses the AB magnitude system, and we conform to that convention
here unless otherwise noted.

With so many sources, it is impossible to distinguish which are the correct
counterparts to the SCUBA detections without some other information, so we now
examine the sub-mm properties of various sub-samples of optical-NIR detected
galaxies.

\subsubsection{Optically Faint Radio Sources (OFRS)}
Observations by \citet{2000AJ....119.2092B} and \cite{2001ApJ...548L.147C}
find that optically faint ($I(\textrm{Vega})>25$) galaxies detected using deep
1.4\,GHz radio observations are coincident with ${\simeq}\,70$ per cent of 
SCUBA sources with fluxes above 5\,mJy.
This population has
been extensively modeled \citep{2002ApJ...570..557C,2003ApJ...585...57C} 
and exploited to obtain
spectroscopic redshifts of the optically detected host galaxies
\citep{2003Natur.422..695C}.

We searched for $I(\textrm{Vega}){>}\,24$ galaxies within 2\arcs\ of a
1.4\,GHz radio detection, and found 17 candidates.
There were an additional 13 radio
sources with no optical counterpart, for a total of 30 optically faint
radio sources. Of these, 9 are within 7\arcs\ of a
SCUBA detected object. These form a subset of the 11 1.4\,GHz detections found
coincident with SCUBA objects without any other selection criteria.
Therefore the optically faint radio sources (OFRS) have a
$9/30 = 30$ per cent success rate in picking out SCUBA sources in the HDF-N,
while the radio sources alone give only $10/135 = 7$ per cent.
Thus OFRS are more effective
at selecting sub-mm bright galaxies than radio alone.  We do not find as high a rate
as some other studies, but it is unclear that these
fractions can be applied directly to other surveys; in the
HDF-N Super-map the sensitivity to sub-mm sources is strongly variable across
the field. If we restrict this analysis to the region of the sub-mm map that
has an RMS noise of 1.5\,mJy or lower, there are 6 out of 13 optically 
faint radio sources coincident with SCUBA detections (46 per cent).

\subsubsection{Galaxies with red optical-NIR colours}
\begin{figure}
\begin{center}
\includegraphics[width=3.0in,angle=0]{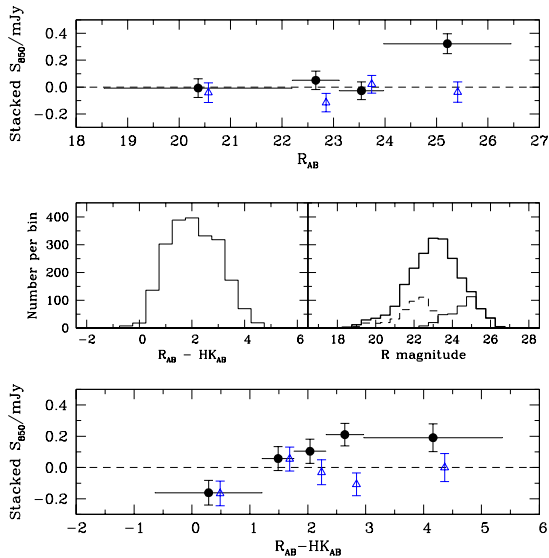}
\caption{The top panel shows the stacked 850\mum\ flux density
as a function of $R$-band magnitude,
showing a marked increase toward optically
fainter objects.  Filled
circles denote the average sub-mm flux density
in each bin, the width of which is given
by the horizontal bars, chosen to ensure that there is the same number of
objects in each bin.  Triangles correspond to the same bins (offset for
clarity), but first removing the known sub-mm sources from the entire
${>}\,3.5\sigma$ SCUBA catalogue.  The central left panel simply shows the
distribution of $R-HK^\prime$ values from the catalogue of \citet{2004AJ....127..180C}.
On the right, we plot the distribution of $R$-band magnitudes (thick line) as
well as the subset of $R$-band fluxes from the reddest (thin line) and bluest
(dashed) bins.  This shows a trend for the redder sources to be fainter, which
means we cannot completely separate colour effects from brightness effects.
The bottom panel plots the stacked 850\mum\ flux density
for 5 bins selected by colour.
The redder objects are statistically detected, but with no detection once the
${>}\,3.5\sigma$ SCUBA sources are removed (triangles).
}
\label{fig:stackero}
\end{center}
\end{figure}

The dust responsible for the extreme IR luminosities that SCUBA detects is also
responsible for reddening the optical-UV spectrum, making such galaxies appear
very red compared with the field population. \citet{2002ApJ...577L..83W}
compared 850\mum\ flux densities against a list of galaxies detected with
$K^\prime({\rm Vega}) < 21.25$, finding a significant trend of increasing
sub-mm flux density with increasing optical-NIR redness.
Fig.~\ref{fig:stackero}
presents the results from a similar analysis we conducted using the HDF-N maps.
In general, the fainter, redder galaxies are more sub-mm bright.  However, we
cannot easily separate the effects of redness from faintness, since they are
correlated (see left part of middle panel of Fig.~\ref{fig:stackero}).  There
is weak evidence that the stacked flux density
does not continue to rise for the most
extremely red sources, though the completeness at these faint flux levels is
lower.  One important thing to note is that the stacked average is dominated
by \textsl{detected} sources, a point also raised in \citet{2004ApJ...605..645W}.

The criterion $R-K>5.3$ (Vega) can be used to define an ERO, motivated by being
the approximate colour of an elliptical galaxy at $z=1$.
Applying a correction to the
$HK^\prime$ magnitudes to convert to $K$ (P.~Capak, private communication), we
calculate that $R - HK^\prime > 3.9$ is the equivalent criterion in AB
magnitudes.  We find 121 EROs within the Super-map.
However, the stacked flux density at their positions is only $0.22\pm0.16\,$mJy,
indicating that EROs themselves are a poor way to pre-select SCUBA galaxies.
This result echoes that of \citet{2002A&A...383..440M}, who targeted a sample
of 27 EROs with SCUBA and detected none down to an RMS of ${\sim}\,2\,$mJy.
Only a single SCUBA source in our catalogue has an ERO within
7\arcs\ (SMMJ123700+620910), and that was already detected in the radio.
However, one has to be careful with direct comparisons, since the near-IR data
we have used are shallower than those available in some other fields, and
it may well be that one finds a higher ERO rate with deeper $K$-band imaging.

\subsubsection{Optical LBG drop-out surveys}
Because of their inferred high star-formation rates and large co-moving number
density, Lyman-Break Galaxies
(LBGs, \citealt{1998ApJ...492..428S}) make an interesting target for sub-mm
studies.  Using reasonable conversion factors based on local standards, the
predicted 850\mum\ flux densities are expected to be on the order of 5\,mJy.
However, sub-mm surveys to date have failed to detect LBGs, except in a 
handful of cases \citep{2000MNRAS.319..318C,2001dmsi.conf...97C,2000MNRAS.318..535P,2003ApJ...582....6W}.  

The HDF-N Super-map covers much more area than previous attempts, and
contains many more known LBGs. \citet{2003ApJ...592..728S} find 132 LBG 
candidates (selected photometrically) in a 73 square arcminute region within
the HDF.   However, the stacked 850\mum\ flux density is $0.03\pm0.10\,$mJy
confirming previous results that LBGs are typically SCUBA-faint. We also 
performed a cross-clustering analysis like that described in the X-ray 
section, but failed to detect any signal. 

In the HDF-N region, only two LBGs seem co-incident with a SCUBA
source. SMMJ123656+621201 (also known as HDF\,850.2) has a pair of possible
LBG counterparts located within 5\arcs.   However, as we discuss in
Section~\ref{sec:counterparts}, this pair may not constitute the correct
ID.  In any case, sub-mm bright galaxies are much rarer than LBGs, with only
1 SCUBA galaxy (with $S_{850}>5\,$mJy)  per 7 arcmin$^2$, compared with roughly
8 LBGs (at $z\sim3$) over the same area.  Hence, even if there is complete
overlap between the populations, number counts alone suggest that the typical
LBG must have an average 850\mum\ flux density of a fraction of a mJy.

\subsubsection{Are optical surveys a good way to identify SCUBA sources?}
\citet{2003ApJ...597..680W} argue that NIR selected sources found in deep
($K_{\rm AB}\sim23$) imaging are just as effective at identifying SCUBA
sources as 1.4\,GHz images (with a $1\sigma$ RMS of $\sim16\,\mu$Jy) in the
CUDSS 14h sub-mm field \citep{1999ApJ...515..518E}.  However, this only
relates to finding a counterpart of a \textsl{known} SCUBA source.  
We have explored various brightness and colour cuts and find that no optical-only 
criteria is successful at \textsl{pre-selecting} SCUBA galaxies.  

Nevertheless, we can examine the optical properties of the radio sources to
see if they have anything in common that can be used to select the counterpart
to a known SCUBA detection. In Table ~\ref{tab:vlaoptical} we present a 
list of the radio detected SCUBA sources (from the $>4\sigma$ catalogue) 
in the HDF-N with optical/NIR sources detected nearby .  
We restrict the list to the
6 radio detected objects that have an optically detected counterpart.
There is one additional source that meets this criterion 
(SMMJ123652+621225), but we exclude it from the list for reasons explained 
later.

We find that of the 6 SCUBA galaxies we are considering, 3 have the
radio source also coincident with the optical source closest to the centroid.
Still, one would be hard-pressed, without the radio data, to
decide between the objects around SMMJ123645+621449.  
In short, though some of the optical-only IDs would have selected the
correct counterpart, there are many cases, especially in crowded fields, where
it becomes too difficult to ascertain with any confidence.  We do note,
however, that in half the cases we would have chosen the correct ID just by picking
the reddest galaxy within 7\arcs.
\begin{table}
\caption{Optical galaxies within 7\arcs\ of radio detected SCUBA galaxies in
the HDF-N.  Here $\theta$ is given in arcseconds, and magnitudes are in the
AB system.  A `Y' in the last column indicates that the galaxy is
radio-detected.}
\label{tab:vlaoptical}
\begin{tabular}{lllll}
\hline
Object Name       & $K$            & $(R-K)$     & $\theta$  & Radio \\\hline\hline  

SMMJ123616+621516 & $22.4\pm0.3$   & $1.5\pm0.3$ & 3.1 &   \\
                  & $23.3\pm0.6$   & $2.1\pm0.6$ & 4.2 & Y \\
SMMJ123618+621552 & $22.8\pm0.4$   & $2.4\pm0.4$ & 3.5 & Y \\   
                  & $>22.1$        & $<3.0$      & 6.6 &   \\
SMMJ123645+621449 & $>22.1$        & $<3.5$      & 1.9 &   \\
                  & $>22.1$        & $<4.1$      & 2.2 & Y \\
                  & $23.2\pm0.4$   & $2.1\pm0.4$ & 2.9 &   \\
                  & $22.8\pm0.3$   & $1.1\pm0.3$ & 3.2 &   \\
                  & $22.8\pm0.3$   & $2.2\pm0.3$ & 5.4 &   \\
SMMJ123650+121316 & $21.4\pm0.2$   & $2.4\pm0.2$ & 0.6 &   \\
                  & $19.9\pm0.1$   & $0.5\pm0.1$ & 5.5 &   \\
                  & $20.4\pm0.1$   & $1.7\pm0.2$ & 6.7 & Y \\
SMMJ123707+621410 & $22.4\pm0.3$   & $3.2\pm0.3$ & 4.3 & Y \\
                  & $22.3\pm0.3$   & $1.6\pm0.3$ & 4.6 &   \\
                  & $>22.1$        & $<3.9$      & 6.7 &   \\
SMMJ123701+621146 & $21.4\pm0.1$   & $3.0\pm0.1$ & 2.6 & Y \\
                  & $>22.1$        & $<4.2$      & 3.8 &   \\
                  & $22.2\pm0.1$   & $2.0\pm0.1$ & 4.9 &   \\
                  & $22.0\pm0.1$   & $0.6\pm0.1$ & 4.9 &   \\
                  & $23.0\pm0.2$   & $1.5\pm0.2$ & 5.9 &   \\
\hline
\end{tabular}
\end{table}

\citet{2003ApJ...597..680W} also claim that clumps of 2 or more nearby EROs
demarcate regions where a sub-mm source can be found.   It is certainly
interesting to see if regions of over-density in one catalogue are good tracers
of objects in another.  Thus, we took the optical
catalogues and counted the number of objects that have at least one neighbour
within 10, 20, and 30 arcseconds of another.  We then assess how many sub-mm 
sources would be detected within the size of the SCUBA array if clumps of 
red galaxies were targeted.

For galaxies with colours $R-HK^\prime>3.9$ (EROs), we find 6 that have at 
least one neighbour within
10\arcs.  The 3 SCUBA pointings required to cover these 6 objects would have
found a single source in our $4\sigma$ sub-mm catalogue.  This is roughly the
same efficiency as simply pointing the telescope at random spots on the sky.
Overall, we can find no cut on the optical catalogues using spatial
correlations, $R-HK^\prime$ colour, or magnitude that would allow us to
pre-select regions where SCUBA sources would be found.

\subsection{\textsl{ISO} mid-IR imaging}
\textsl{ISO} observed a roughly $\sim4\arcm\times5\arcm$ area around the HDF-N
at both 7 and 15\mum.   The data were reduced separately by three different
groups \citep{1997MNRAS.289..465G,1999A&A...342..313A,1999A&A...342..363D}, and
here we use the 15\mum\ catalogue found in \citet{1999A&A...342..313A}.  
They present a list of 49 objects with high SNR, and an
additional 50 with lower confidence, covering a total area of 24 arcmin$^2$.
The positions are accurate to within 3\arcs.

Concentrating first on the 49 secure \textsl{ISO} 15\mum\ detections, we find a
stacked 850\mum\ flux density of $0.65\pm0.09\,$mJy, which drops to
$0.16\pm0.09\,$mJy after removing sections around SCUBA sources.
The list of 50 less secure detections gives similar results, and the overall
(99 sources) stacked flux density is
significantly detected at $0.41\pm0.06\,$mJy.  The brightest third of the
\textsl{ISO} sources ($S(15$\mum$)>150\mu$Jy) have the strongest 
stacked flux density, with an average value of $0.86\pm0.13\,$mJy.
We later conclude that only 1 object from our catalogue has a 15\mum\
counterpart, despite the low probability that 4 of 13 SCUBA galaxies would
have an {\sl ISO\/} source within 7\arcs\ by chance (see Table~\ref{tab:pstat}).
Of the other SCUBA sources with nearby {\sl ISO\/} sources, one is identified
with an unrelated radio source and the remaining are inconclusive.

\section{Counterparts of each SCUBA object}
\label{sec:counterparts}

\subsection{General criteria for counterpart identification}
We now come to a detailed description of the counterparts of the SCUBA
detected objects.
Of all the correlations presented thus far, the 1.4\,GHz--850\mum\ overlap is
by far the strongest.  Thus in this section there will naturally be an emphasis
on the radio sources.  Based on the strength of this correlation, we will not
only consider the sources formally detected at $5\sigma$ in the 1.4\,GHz
map presented by \citet{2000ApJ...533..611R}, but fainter ones as well.
\citet{2000ApJ...533..611R} find that the 1.4\,GHz differential number counts
are well fit by a power law of index $\gamma=-2.4\pm0.1$.  Assuming uniform
noise, the number of sources brighter than an $n\sigma$ noise limit goes as
$N(>5\sigma)(n/5)^{1+\gamma}$.  Thus for a $3\sigma$ cut, there are about
twice as many sources.  This is still small
enough that chance overlap with a large number of SCUBA detected galaxies is
negligible.  There is no question that a few of these fainter sources will be
spurious however, so we lend them weight only if there is an object detected
in the optical, NIR, or X-ray bands at the same position.

We will not assign a secure ID to SCUBA sources unless they are detected at
$>3\sigma$ in the radio, but will offer tentative IDs for sources that are
faint and red, or are the only object detected within the error circle.
Together, this leads to a reasonable recipe for identifying counterparts for
the SCUBA objects:
\begin{enumerate}
\item Strongly detected ($>5\sigma$) 1.4\,GHz sources within 7\arcs\ of the
SCUBA centroid will be deemed the host of the sub-mm emission.
Optical, NIR, and X-ray properties will be based on the object that is
coincident with the radio position (if any). 
\item Weakly detected ($3-5\sigma)$ 1.4\,GHz sources within 7\arcs, with a
counterpart in another band that has similar or better resolution, will be
chosen as the identification with the sub-mm source.
\item A detection in any band, within the search radius, will be flagged as a
possible identification if and only if no other candidate in the vicinity is
detected.
\end{enumerate}

A description of each source, and images of each are presented in the appendices
(Figs.~\ref{fig:lw4postage}--\ref{fig:sw4postage}). We summarize the counterparts
in  Table~\ref{tab:multisum}. We suggest 10 secure identifications in the 
$>4\sigma$ SCUBA source list, and a
further 2 tentative IDs, as well as 5 secure IDs for the 3.5--$4.0\sigma$
sources and 1 tentative for the $450\mu$m sources.  Although in each case the radio
galaxy is used for the secure counterpart, our selection criterion apparently
fails for the one special source we have alluded to throughout this paper.

\subsection{The curious case of SMMJ123652+621225} 
\label{sec:hdf850p1}
The radio detected \chandra\ galaxy 5.7\arcs\ to the South-west of this SCUBA source
seemingly obvious choice.  However, deeper observations have revealed that
this is not the case. This
sub-mm source, usually called HDF\,850.1, has been the subject of intense
scrutiny since its discovery in the initial deep sub-mm survey of the HDF by
\citet{1998Natur.394..241H}. After years of effort, 
\citet{2004MNRAS.350..769D} now claim that the
counterpart for 850.1 has been determined.

Using the IRAM interferometer, the sub-mm flux was 
resolved with a positional
uncertainty of $\sim0.3$\arcs\ \citep{1999A&A...347..809D}.  However, it was
not found coincident with the VLA source mentioned above, but rather a new 
and very faint radio source found by co-adding VLA and MERLIN data. 

It cannot be stressed how difficult obtaining this (tentative)
counterpart has been.  Fortunately,
the majority of SCUBA objects do not appear to
suffer from being in a such a complicated system.  Nevertheless, experience
with this object
only highlights the need for better angular resolution and
sensitivity for sub-mm observations, as it calls into question the validity of
{\em any\/} counterpart.

\begin{table*}
\caption{Multi-wavelength summary of SCUBA detections.  We give the source
name, counterpart position and distance from the SCUBA centroid, radio, sub-mm,
mid-IR, and optical/near-IR fluxes (or $3\sigma$ upper limits).  Optical
magnitudes are in the AB system.  The X-ray fluxes are presented separately in
Table~\ref{tab:xrayprops}.  The symbols in the `Bands' column are as follows:
R means detected in the 1.4\,GHz radio map; 8 is the 8.5\,GHz radio; X is
X-ray; I is Mid-IR (\textsl{ISO}); and O means an optical counterpart.  The
`Comments' column indicates whether we regard the ID as secure or tentative.
Gaps in the table mean that imaging was not done at that position.  
For consistency, we provide the data on the radio source
near HDF850.1, although this may not be the correct ID.
}
\scriptsize
\label{tab:multisum}
\begin{tabular}{llllllllllll}
\hline
SMM ID         &\multicolumn{2}{c}{J2000}       & $\theta$ & $S_{\rm 1.4GHz}$ &$S_{\rm 8.5GHz}$ & $S_{\rm 850\mu m}$ & $S_{\rm 450\mu m}$ & $S_{\rm 15\mu m}$ & Optical            & Bands & Comments \\
               & 12h$+$          & $62^\circ+$  & (\arcs\,)& ($\mu$Jy)        &($\mu$Jy)        & (mJy)              & (mJy)              & ($\mu$Jy)         & $R,R-HK^\prime$           &       &    \\\hline\hline
\multicolumn{12}{c}{850\mum\ detections $>4\sigma$}\\\hline
J123607+621143 &           &          &          & $<45$           & $<9$             & $15.2\pm3.8$       & $<97$              &                   &                    &       & \\ %R48067
J123608+621249 & 36:08.530 & 12:49.61 & 0.4      & $<45$           & $<9$             & $16.0\pm3.7$       & $<95$              &                   & $25.2\pm0.1, <3.1$ & O     & Tentative\\ %R47569
J123616+621516 & 36:16.148 & 15:13.67 & 4.4      & $53.9\pm8.4$    & $<9$             & $6.3\pm0.9$        & $<35$              &                   & $25.4\pm0.1,  2.1$ & RXO   & Secure \\%R46504
J123618+621007 &           &          &          & $<45$           & $<9$             & $6.6\pm1.5$        & $<66$              &                   &                    &       & \\%R30091
J123618+621552 & 36:18.328 & 15:50.48 & 3.5      & $151\pm11$      & $<9$             & $7.2\pm0.9$        & $<34$              &                   & $25.2\pm0.1,  2.4$ & RO    & Secure\\ %R46195
J123621+621252 &           &          &          & $<45$           & $<9$             & $12.1\pm2.6$       & $<85$              &                   &                    &       & \\%R47535
J123621+621710 & 36:21.272 & 17:08.40 & 2.5      & $148\pm11$      & $<9$             & $8.8\pm1.5$        & $<72$              &                   &                    & R    & Secure\\%R45640
J123622+621616 & 36:22.625 & 16:21.28 & 4.4      & $<45$           & $<9$             & $8.6\pm1.0$        & $51\pm17$          &                   & $24.4\pm0.1, <2.3$ & O     & Tentative (pair)\\%R46009
J123634+621407 &           &          &          & $<45$           & $<9$             & $11.2\pm1.6$       & $<67$              & $<200$            &                    &       & \\
J123637+621155 & 36:37.565 & 11:56.32 & 4.0      & $40\pm9$        & $<9$             & $7.0\pm0.8$        & $<46$              & $<200$            &                    & R     & Secure\\
J123645+621449 & 36:46.049 & 14:48.69 & 2.2      & $124.0\pm9.8$   & $24.7\pm4.5$     & $8.5\pm1.3$        & $<47$              & $<200$            & $26.2\pm0.2, <4.1$ & R8XO  & Secure\\%R46661
J123650+621316 & 36:49.708 & 13:12.78 & 7.0      & $49.2\pm7.9$    & $<9$             & $2.0\pm0.4$        & $<11$              & $<200$            & $22.1\pm0.1, 1.7$  & RXOI  & Secure\\%R47395
J123652+621225 & 36:51.760 & 12:21.30 & 5.7      & $49.3\pm7.9$    & $16.8\pm2.3$     & $5.9\pm0.3$        & $<12$              & $<200$            & $22.0\pm0.1, 1.6$          &       & (HDF\,850.1)\\%R47767&
J123656+621201 & 36:56.605 & 12:07.62 & 5.8      & $46.2\pm7.9$    & $<9$             & $3.7\pm0.4$        & $<16$              & $<200$            &                    & R     & Secure\\
J123700+620910 & 37:00.256 & 09:09.75 & 1.5      & $324\pm18$      & $66.7\pm13.7$    & $8.6\pm2.1$        & $<85$              &                   &                    & R8    & Secure\\
J123701+621146 & 37:01.574 & 11:46.62 & 2.3      & $128.0\pm9.9$   & $29.5\pm2.8$     & $4.0\pm0.8$        & $<27$              & $<200$            & $25.3\pm0.1,  3.9$ & R8XO  & Secure (ERO)\\%Z1896
J123702+621301 &           &          &          & $<45$           & $<9$             & $3.4\pm0.6$        & $<25$              & $<200$            &                    &       & \\
J123707+621410 & 37:07.208 & 14:08.08 & 4.4      & $45.3\pm7.9$    & $29.0\pm6.1$     & $9.9\pm2.5$        & $<85$	        &                   & $25.6\pm0.1,  3.3$ & R8XO  & Secure \\%R43855
J123713+621204 &           &          &          & $<45$           & $<9$             & $6.1\pm1.4$        & $<40$              &                   &                    &       & \\\hline
\multicolumn{12}{c}{850\mum\ detections $3.5$--$4.0\sigma$}\\\hline
J123607+621019 & 36:06.850 & 10:21.38 & 3.2      & $74.4\pm9.0$    & $<9$             & $13.5\pm3.7$       & $<97$              &                   & $24.4\pm0.1,  2.8$ & RXO   & Secure\\%R29898
J123608+621431 & 36:08.592 & 14:35.77 & 4.1      & $68.9\pm8.8$    & $<9$             & $6.1\pm1.7$        & $<60$	        &                   &                    & RX    & Secure\\%R46768
J123611+621213 &           &          &          & $<45$           & $<9$             & $12.8\pm3.4$       & $<90$              &                   &                    &       &  \\
J123628+621046 & 36:29.134 & 10:45.79 & 3.4      & $81.4\pm8.7$    & $<9$             & $4.4\pm1.2$        & $<55$	        &                   & $24.0\pm0.1,  3.6$ & RXO   & Secure\\%R29754
J123635+621237 & 36:34.515 & 12:41.01 & 7.6      & $230.0\pm14$    & $52.6\pm4.7$     & $3.0\pm0.8$        & $<41$              & $363_{-38}^{+79}$ & $23.4\pm0.1,  2.6$ & R8XOI & Secure\\%R47602
J123636+620658 &           &          &          & $<45$           & $<9$             & $22.1\pm5.6$       & $<155$	        &                   &                    &       & \\
J123647+621840 &           &          &          & $<45$           &                  & $19.5\pm5.4$       & $<154$	        &                   &                    &       & \\
J123652+621352 & 36:52.76  & 13:54.1  & 2.0      & $<45$           & $7.8\pm2.8$      & $1.8\pm0.4$        & $<16$	        & $<200$            & $22.3\pm0.1,  0.9$ & 8XO   & Secure\\%R43949
J123653+621119 &           &          &          & $<45$           & $<9$             & $2.8\pm0.8$        & $<40$              & $<200$            &                    &       & \\
J123659+621452 &           &          &          & $<45$           & $<9$             & $5.2\pm1.4$        & $<72$	        & $<200$            &                    &       & \\
J123706+621849 &           &          &          & $<45$           &                  & $21.6\pm5.8$       & $<178$	        &                   &                    &       & \\
J123719+621107 &           &          &          & $<45$           & $<9$             & $7.2\pm2.0$        & $<55$              &                   &                    &       & \\
J123730+621055 &           &          &          & $<45$           & $<9$             & $13.3\pm3.6$       & $<98$              &                   &                    &       & \\
J123730+621855 &           &          &          & $<45$           & $<9$             & $27.1\pm7.6$       & $<286$	        &                   &                    &       & \\
J123741+621225 &           &          &          & $<45$           &                  & $23.7\pm6.1$       & $<185$	        &                   &                    &       & \\\hline
\multicolumn{12}{c}{450\mum\ detections $>4\sigma$}\\\hline
J123619+621127 & 36:19.370 & 11:25.6  &   2.0    & $<45$           & $<9$             & $<5.8$             & $110\pm26$         &                   & $24.0\pm0.1,  1.4$ &       & Tentative\\ %R48239
J123632+621542 &           &          &          & $<45$           & $<9$             & $<5.9$             & $105\pm25$         &                   &                    &       & \\
J123702+621009 &           &          &          & $<45$           & $<9$             & $<5.2$             & $120\pm27$         &                   &                    &       & \\
J123727+621042 &           &          &          & $<45$           & $<9$             & $<10.4$            & $220\pm42$         &                   &                    &       & \\
J123743+621609 &           &          &          & $<45$           & $<9$             & $<24.0$            & $300\pm72$         &                   &                    &       & \\\hline

\normalsize
\end{tabular}
\end{table*}

\subsection{Available redshift information for the SCUBA sources in the HDF-N}
\label{sec:redshifts}
Determining redshifts of the SCUBA population is critical in order to place 
these systems in their correct cosmological context, but this has proven
difficult observationally. 

Extensive 
spectroscopic redshift campaigns have been performed in the HDF-N
over the past several years
\citep{2000ApJ...538...29C,2000AJ....119.2092B,2002AJ....124.1839B}.
An accounting of available spectroscopic and photometric redshifts 
was recently presented in \citet{cowiez} and \citet{2004AJ....127.3121W}.
Unfortunately the overlap with SCUBA counterparts is low, so we are forced 
to use FIR photometric redshift estimators as well.  In 
Table~\ref{tab:hdfredshift} we provide estimates based on the sub-mm only
data, as well as those derived using the Carilli-Yun method
\citep[][hereafter called the `CY estimator']{1999ApJ...513L..13C,2000ApJ...530..618C} which also employs radio information.  Since none
of the sources are detected at 450\mum, the sub-mm limits are essentially
useless, but even with high signal-to-noise detections at 450 and 850\mum,
estimating redshifts using FIR SEDs is difficult \citep{2003MNRAS.338..733B}.
 The CY-estimator is much more constraining.

For the few sources that do have spectroscopic
redshifts, the photometric redshifts seem reasonable, although the agreement
is far from perfect.  The most obvious exceptions are SMMJ123607+621019 and
SMMJ123608+621431, which prefer larger redshifts than found by optically
based 
measurements.  We already noted that the field around both are quite complex,
and lensing may be a factor.  However, without confirmation from a CO line
detection, higher resolution sub-mm data, or further SED information, we are
unable to determine whether these are unusual SCUBA galaxies or the wrong
identification.

\begin{table}
\caption[Redshift summary of HDF-N sub-mm sources.]{Redshift summary of HDF-N
sub-mm sources. CY estimates are based on the most
likely 1.4\,GHz counterpart in the list, or the lower limit in the case of a
non-detection.  Since the CY-estimator is not effective past $z\sim3$, we cap
the lower limits there.  The estimates based on ratios between 850\mum\ and
450\mum\ flux densities are listed as well, but only for those cases with
$S_{450}/S_{850}< 6.7$, which is the ratio of Arp\,220 at
redshift zero.  $z_{\rm optical}$ is the redshift of the counterpart based on
spectroscopy or optical photometry. A question mark is placed in that
column if the counterpart has no redshift estimate yet.  Blank entries in this
column denote sources where we were unable to determine a counterpart at all.
}
\label{tab:hdfredshift}
\begin{tabular}{llll}
\hline
ID                & $z_{\rm CY}$            &$z_{\rm 450/850}$  &$z_{\rm optical}$\\\hline\hline
\multicolumn{4}{c}{850\mum\ detections $>4\sigma$}\\\hline
     SMMJ123607+621143  &$>3.0$                   &$>0.4$  &                  \\
     SMMJ123608+621249  &$>3.0$                   &$>0.7$  & ?                \\
     SMMJ123616+621516  &$2.5_{-0.8}^{+1.7}$      &$>1.1$  & 2.06$^{a,b}$     \\
     SMMJ123618+621007  &$>2.9$                   &        &                  \\
     SMMJ123618+621552  &$1.6_{-0.4}^{+0.8}$      &$>1.8$  & ?                \\
     SMMJ123621+621252  &$>3.0$                   &        &                  \\
     SMMJ123621+621710  &$1.8_{-0.5}^{+1.0}$      &        & ?                \\
     SMMJ123622+621616  &$>3.0$                   &$>0.4$  & ?                \\
     SMMJ123634+621407  &$>3.0$                   &$>0.3$  &                  \\   
     SMMJ123637+621155  &$3.2_{-1.2}^{+2.4}$      &$>0.2$  & ?                \\
     SMMJ123645+621449  &$1.9_{-0.5}^{+1.0}$      &$>1.1$  & ?                \\
     SMMJ123650+621316  &$1.5_{-0.5}^{+0.8}$      &$>1.1$  & 0.475$^{c}$      \\
     SMMJ123652+621225  &$>2.7$                   &$>4.9$  &                  \\
     SMMJ123656+621201  &$2.0_{-0.6}^{+1.3}$      &$>2.2$  & ?                \\
     SMMJ123700+620910  &$1.3_{-0.4}^{+0.5}$      &        & ?                \\ 
     SMMJ123701+621146  &$1.4_{-0.4}^{+0.6}$      &        & 1.52$^{a,b}$     \\
     SMMJ123702+621301  &$>2.0$                   &        &                  \\
     SMMJ123707+621410  &$3.7_{-1.5}^{+2.8}$      &        & ?                \\
     SMMJ123713+621204  &$>2.7$                   &$>0.2$  &                  \\\hline
\multicolumn{4}{c}{Additional 850\mum\ detections $>3.5\sigma$}\\\hline
     SMMJ123607+621019  &$3.3_{-1.2}^{+2.5}$      &        & 0.47$^{a,b}$     \\
     SMMJ123608+621431  &$2.1_{-0.7}^{+1.4}$      &        & 0.472$^{c}$      \\
     SMMJ123611+621213  &$>3.0$                   &        &                  \\
     SMMJ123628+621046  &$1.7_{-0.5}^{+0.9}$      &        & 1.013$^d$        \\
     SMMJ123635+621237  &$1.0_{-0.4}^{+0.4}$      &        & 1.219$^d$        \\
     SMMJ123636+620658  &$>3.0$                   &        &                  \\
     SMMJ123647+621840  &$>3.0$                   &        &                  \\
     SMMJ123652+621352  &$>1.5$                   &        & 1.355$^d$        \\
     SMMJ123653+621119  &$>1.8$                   &        &                  \\
     SMMJ123659+621452  &$>2.4$                   &        &                  \\
     SMMJ123706+621849  &$>3.0$                   &        &                  \\
     SMMJ123719+621107  &$>3.0$                   &        &                  \\
     SMMJ123730+621055  &$>3.0$                   &        &                  \\
     SMMJ123730+621855  &$>3.0$                   &        &                  \\
     SMMJ123741+621225  &$>3.0$                   &        &                  \\\hline
\end{tabular}
\medskip
\\
$^a$\,From \citet{2002AJ....124.1839B}.\\
$^b$\,Redshift determined photometrically from optical counterpart.\\
$^c$\,From \citet{cowiez}.\\
$^d$\,From \citet{2000AJ....119.2092B}.\\
\end{table}

\section{Caveats to source identifications}
It cannot be stressed strongly enough that the $\mu$Jy 1.4\,GHz radio detections
have a high degree of overlap with SCUBA sources, and that they currently
constitute the best way to find counterparts to the sub-mm detected objects.
However, there are some cases where the ID still remains ambiguous.  HDF\,850.1
(SMMJ123652+621225) is a particularly pathological example of how the nearby
VLA source may \textsl{not} be the correct ID.  SMMJ123645+621449 is a case
where 2 radio sources are present near the sub-mm source, and it is difficult to
choose which might be the correct ID.  Thus the radio-detected sub-mm galaxies
that are shaping our understanding of the entire SCUBA population are by no
means immune to selection biases.  However, it is encouraging that follow-up
IRAM observations are detecting CO emission at the redshifts determined via
optical spectroscopy 
\citep{1999ApJ...514L..13F,2003ApJ...599..839G,2003ApJ...597L.113N}.

One also has to be aware of two other complications with the identification
process -- clustering and lensing.  Since sub-mm galaxies are believed to be
associated with merging systems, then it is likely that in many cases the
sub-mm bright galaxy will be physically associated {\it but distinct from\/}
one or more other galaxies which may be sub-mm faint.  So one has to be careful
when investigating the SEDs, that one is not combining data from multiple
objects with very different spectral properties.

Lensing is a further complication, which we know is an important issue for
some sub-mm sources in rich cluster fields \citep{2004MNRAS.349.1211K,borysms}
 and has been suggested for a couple of sources in the HDF region
(SMMJ123652+621225 and SMMJ123637+621155).  Lensing may
make a sub-mm source apparently associated with a galaxy which actually in the foreground.

\section{Summary and Future Work}
Concentrating on the more secure (${>}\,4\sigma$) SCUBA sources, we find that
10 out of 19 have a VLA 1.4\,GHz radio counterpart, and at least one more has
radio detected at a fainter
level in a combined map (J123652+6211225, a.k.a. HDF\,850.1).   This is in line
with results from \citet{2002MNRAS.337....1I}, \citet{2002ApJ...570..557C},
\citet{2000AJ....119.2092B}, and \citet{greve}.   
This strong overlap, along with the well
determined radio positions,  has recently been exploited to obtain redshifts
for a large fraction of this subset of sources \citep{2003Natur.422..695C}.
However, significant improvements in radio sensitivity are required in order to
detect the fainter SCUBA sources, particularly if some of them lie at higher
redshifts.  Because of this, one cannot simply accept the nearest detected
radio source as the correct ID.  For example, we argue (see Appendix)
in the case of SMMJ123622+621618 that
the radio source 13\arcs\ away is \textsl{not} responsible for the sub-mm
emission (as reported in \citealt{2000AJ....119.2092B}).  Radio sources on
their own are not a good way to pre-select SCUBA galaxies, but we note that the
sub-sample of radio detections with $I>24$ have at least a roughly
30 per cent
detection rate in the SCUBA map of HDF-N.  This is a lower limit, since much of
the sub-mm map is simply too insensitive to sources with $S_{850}\sim 5$\,mJy,
which is the flux density level where \citet{2002ApJ...570..557C} find a
much higher (roughly 70 per cent) success rate.

Although X-ray observations help us understand the nature of the SCUBA 
detected galaxies, they are not useful in the counterpart ID process in the 
HDF region.  The ID rate is such
that the VLA can detect, in a tenth of the time, all the SCUBA galaxies
detected by \chandra.

There is weak evidence suggesting that the reddest object in the vicinity
of a SCUBA source is correct optical counterpart, but we found no compelling
evidence that optical properties alone aid in determining counterparts.

Despite the wealth of deep multi-wavelength data in this part of the sky,
almost half of our sources have an undetermined counterpart,
clearly demonstrating how difficult it is to
make secure identifications.  Nevertheless, continued sub-mm imaging in concert
with deep \hst\ and \sirtf\ data in the region should allow us to characterise
the rest-frame UV-NIR spectra of at least the radio-identified galaxies.
Confidence in the radio-unidentified half needs to wait for higher resolution 
sub-mm images from ALMA.  But in the meantime there is a realistic hope of being able to
obtain a full accounting of the sub-mm galaxies in this one small field, and
in particular to find out the nature of the currently unidentified half.

\section*{Acknowledgments}
We would like to thank Peter Capak, Chuck Steidel, Mark Dickinson, Seb Oliver,
and Tracy Webb for useful discussions. CB also thanks Ian Smail and Andrew 
Blain for useful advice on the content and presentation of this paper.  
We also appreciate advice from the referee which helped streamline the text.
This work was supported in part by the Natural Sciences and Engineering 
Research Council of Canada, as well as by a grant from NASA administered by 
the American Astronomical Society.
The James Clerk Maxwell Telescope is operated by The Joint Astronomy Center on
behalf of the Particle Physics and Astronomy Research Council of the United
Kingdom, the Netherlands Organisation for Scientific Research, and the National
Research Council of Canada. Much of the data for this paper was obtained via
the Canadian Astronomy Data Centre, which is operated by the Herzberg Institute
of Astrophysics, National Research Council of Canada.

\appendix
\section{Multi-wavelength description of sub-mm detected sources in GOODS-N}
\subsection{850$\bmath{\mu}$m sources from the primary catalogue}
\begin{figure*}
\includegraphics[width=7in,angle=0]{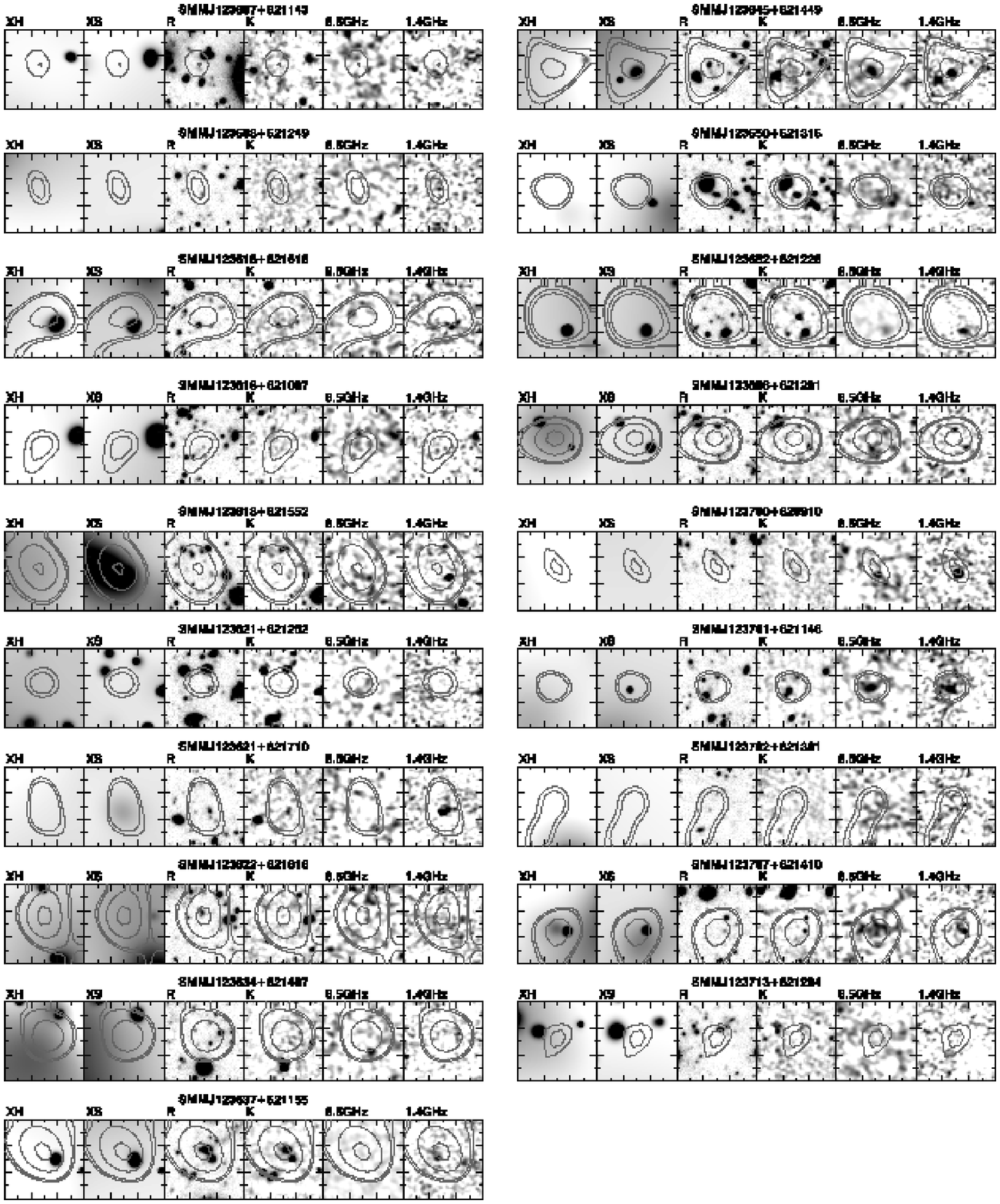}
\caption{Postage stamps of the $>4\sigma$ 850\mum\ SCUBA sources showing: XH
(hard X-ray); XS (soft X-ray); $R$ and $HK^\prime$ (ground based optical
imaging); $8.5\,$GHz and $1.4\,$GHz (VLA radio).  These data
are described in Section~\ref{sec:catdescriptions}.  We show 30\arcs\ on a side
negative images at each waveband, with North and East running towards the top
and left of the page, respectively.
3.5, 4, 6, and $8\sigma$ SCUBA 850\mum\ contours are over-plotted on each.}
\label{fig:lw4postage}
\end{figure*}

Here we describe the 19 objects in the
$4\sigma$ HDF-N SCUBA
catalogue.  We name the objects according to their positions and using the
`SMMJ' prefix, which has become conventional.  In square brackets we use the letters
J, S, and P to indicate that the data were collected by jiggle-map, scan-map,
or photometry observations, respectively.  All optical magnitudes are given in
the AB system.  The 19 strips of postage stamps are shown in
Fig.~\ref{fig:lw4postage}, and some of the collected photometry can be found
in Tables~\ref{tab:xrayprops} and \ref{tab:multisum}.

\noindent\textbf{SMMJ123607+621143 [S]:} A weak radio source detected by WSRT
lies 8\arcs\ to the East. An inspection of the VLA map shows a $4\sigma$ peak
near this position, although it is 2\arcs\ even further East from the SCUBA
galaxy (but consistent with the offset between the catalogues as discussed
previously). Interestingly, the weak VLA source is coincident with a red
($R-HK^\prime=3.3)$ galaxy.  Given the rarity of such sources, it is tempting
to associate the sub-mm galaxy with it, despite the offset. However, this is
well outside our 7\arcs\ search radius, so we dismiss it.  There are 5 optically 
detected objects within the search area but with no further information we cannot 
assign a robust counterpart to this source.

\noindent\textbf{SMMJ123608+621249 [S]:} This source lies directly on top of a
very faint object ($R=25.2\pm0.1$). There are no other optical counterparts
within 7\arcs, and no radio or X-ray objects anywhere near the SCUBA source,
thus suggesting this as the tentative ID.

\noindent\textbf{SMMJ123616+621516 [JS]:} 2 radio sources are found within
7\arcs, one from the WSRT and one from the VLA.  These are not spatially
coincident, but given that the fluxes are comparable, and the WSRT position is
West of the VLA source (again consistent with the offset between the
catalogues), they are likely to be the same source. The optical counterpart is
fairly faint, and appears to have a faint red companion 1\arcs\
to the North-west.  Coincident with the radio ID is a blended \chandra\ source.
Note that this object is discussed in some detail in
\citet{2003AJ....125..383A}. They describe the second X-ray source as invisible
in the radio and optical, and though not formally detected, the postage stamps
do suggest that the second object is red and radio faint. We choose the radio
detected X-ray source as the ID, though point out that the sub-mm flux may be
due to the much fainter optically detected galaxy, or a combination of both. 

\noindent\textbf{SMMJ123618+621007 [JS]:} The VLA source (detected in X-ray and
optical images) 13\arcs\ away is too far to comfortably call it
the counterpart.  Within 20\arcs\ there are 3 LBGs, but the closest is
8.3\arcs\ away. Since we find 2 optically detected galaxies  within 6\arcs,
we cannot assign an ID.  We note an even fainter galaxy which, although not in 
the $>5\sigma$ optical catalogue, is seen in the
optical thumb-nails and seems to lie directly on the SCUBA centroid.

\noindent\textbf{SMMJ123618+621552 [JS]:} A strong 1.4\,GHz VLA source
($151\,\mu$Jy) with a very steep spectrum lies 3.5\arcs\ to the South-west.  An
8.5\,GHz source is also seen there, though not formally listed in the
catalogue.  We consider this radio source to be the counterpart.  The only
optically detected galaxy in the search region is a faint optical source
2\arcs\ North of the VLA position, but a fainter (and formally undetected)
$HK^\prime$ source is coincident with the radio. 

\noindent\textbf{SMMJ123621+621252 [JS]:} Two optically detected galaxies are
present within the search area, meaning we cannot assign a robust ID. Outside
of the 7\arcs\ search radius, the field is rather dense with optically detected
galaxies.  There are also 7 \chandra\ sources in total within 18\arcs\ of the
SCUBA source.

\noindent\textbf{SMMJ123621+621710 [JS]:} An extended radio source is
coincident with the SCUBA position.  Its North-western end is coincident with 
an optically detected galaxy that has an elongated morphology
perpendicular to the radio extension.
Although not listed in the 2\,Ms
\chandra\ catalogue, there is extended soft X-ray emission clearly seen.
We use this VLA source as the SCUBA counterpart.

\noindent\textbf{SMMJ123622+621616 [JS]:} \citet{2000AJ....119.2092B}
associates this object with the radio source 13\arcs\ to the North.  This is
well outside our search radius however, and we do not assign it as the
correct ID. \citet{2003AJ....125..383A} use this object as a possible example
of a Compton-thick AGN, but they noted the inferred redshift
($z=0.46^{+0.03}_{-0.02}$), based on a $2.3\sigma$ iron line in the X-ray
spectrum, was considerably different than that obtained via fits to the
FIR-radio SED of the galaxy ($z=2.4^{+1.1}_{-0.8}$).  In the optical
catalogues, there is a pair of galaxies separated by less than 1\arcs\ and
3 other objects within the 7\arcs\ search radius.

\noindent\textbf{SMMJ123634+621407 [JS]:}  Although the field is dense in
optically detected sources, one unique object in the vicinity is an X-ray
detected LBG with $z=3.408$.  The 11.4\arcs\ offset is rather large, especially
considering the 850\mum\ flux density is well above the confusion level and the
SNR is high.  The ISO 15\mum\ detection 5.3\arcs\ to the South is the more
appealing counterpart, and there is an 8.5\,GHz object coincident with the ISO
centroid.  However it is not listed in the 8.5\,GHz catalogue, and no 1.4\,GHz
or optical flux is detected here. We conclude that there is no robust
counterpart.
%
% figure page 2

\noindent\textbf{SMMJ123637+621155 [JSP]:} A $4\sigma$ radio detection by the
WSRT (but not VLA) does not appear coincident with any other object, but the
VLA map does reveal radio flux, though not quite at the same position.
Seemingly unrelated are two optical candidates, which both have spectroscopic
redshifts, and lie within 4\arcs\ of the SCUBA source.  The first is a
$z=0.779$ galaxy with $R=22.5, R-HK^\prime=1.6$.  The second is a $z=0.557$
galaxy with $R=22.4, R-HK^\prime=1.6$, which also happens to line up with a
soft X-ray source.  At these redshifts it is
difficult to understand the lack of a radio detection if one of these two
objects were the SCUBA counterpart. As explained in
Section~\ref{sec:chandragood}, we suggest that these optically detected
galaxies are lensing the SCUBA source, and that the radio object is the true ID. 

\noindent\textbf{SMMJ123645+621449 [SP]:} There are two 1.4\,GHz VLA sources
coincident with \chandra\ detections near the SCUBA source, and we choose the
one closest to the SCUBA centroid (also the brighter of the two). This object
is also detected at 8.5\,GHz.  The other radio source is
detected by ISO.  Redshifts would be useful for determining if the radio pair
is indeed spatially related.

\noindent\textbf{SMMJ123650+621316 [JSP]:} The counterpart that meets our
criterion is detected in most
wave-bands -- both radio frequencies, 15\mum, X-ray, and
optical, and also has a redshift of $0.475$.   The faintness of
this sub-mm source suggests that confusion noise may be a particular problem,
but if the radio flux is coming from the SCUBA source, then the radio/sub-mm
flux density ratio is a factor of 10 too low for this redshift.  
Note that this SCUBA source is considered as two distinct
objects (HDF\,850.4/HDF\,850.5) by \citet{2003MNRAS.344..887S}, but the
combined sub-mm data that make up the Super-map yield only a single source.
Since its profile is quite Gaussian, we do not treat the possibility of it
being two sources.

\noindent\textbf{SMMJ123652+621225 [JSP]:} This object (a.k.a.~HDF850.1) is
discussed at length in Section \ref{sec:hdf850p1}.

\noindent\textbf{SMMJ123656+621201 [JSP]:} This SCUBA object is
commonly called HDF\,850.2, and at only 3.7\,mJy, has a position that may
be affected by confusion noise.  The faint radio source 5.8\arcs\ to
the North seems like the correct ID, but has no optical counterpart. 
We note that there is a pair of LBGs just
to the South-west, one of which is \chandra\ detected and has radio emission 
just below the VLA catalogues' threshold. The LBG was discussed by
\citet{2002ApJ...576..625N}, who noted that if at redshift 3, this galaxy is
almost certainly an AGN, given its X-ray luminosity.  If at the same redshift,
the small separation of the LBGs suggests they are interacting, which makes
it an appealing choice for the SCUBA counterpart. Another mark in
its favour is that the sub-mm flux density is very small, which would be in
line with the rare sub-mm detections of LBGs. Although the LBGs are the 
more 
interesting case, our counterpart criterion favours
the faint radio detected object.

\noindent\textbf{SMMJ123700+620910 [JS]:}  The only obvious contender is an
optically invisible radio source (detected by the VLA at both frequencies) only
a few arcseconds away from the SCUBA position.

\noindent\textbf{SMMJ123701+621146 [JS]:}  Directly on top of the SCUBA
centroid is a striking radio and X-ray detected ERO. The object 2\arcs\ to the
South-east of the ERO has a redshift of 0.884 and an \textsl{ISO} 15\mum\
detection.  However, it is possible that the ISO flux is actually coming from
the ERO.  Given the proximity of the source, the ERO may well be lensed by the
foreground ($z=0.884$) galaxy.

\noindent\textbf{SMMJ123702+621301 [JSP]:}  The elongated contours suggest 
the blending of 2 nearby sources.
The lower part does seem to be centred on an optically detected
galaxy.  However, we find no convincing counterpart to the SCUBA source.

\noindent\textbf{SMMJ123707+621410 [JSP]:} The correct identification seems
highly likely to be the optically red radio and X-ray detected object right 
on top of the SCUBA source.

\noindent\textbf{SMMJ123713+621204 [JS]:} There are no radio, red objects, or
X-ray sources nearby that are convincing enough to call a counterpart. Two galaxies 
in the $R$-selected catalogue fall within
7\arcs, but we are unable to determine which (if either) is the correct ID.

\subsection{850$\bmath{\mu}$m sources from the supplementary catalogue}
\begin{figure*}
\includegraphics[width=7in,angle=0]{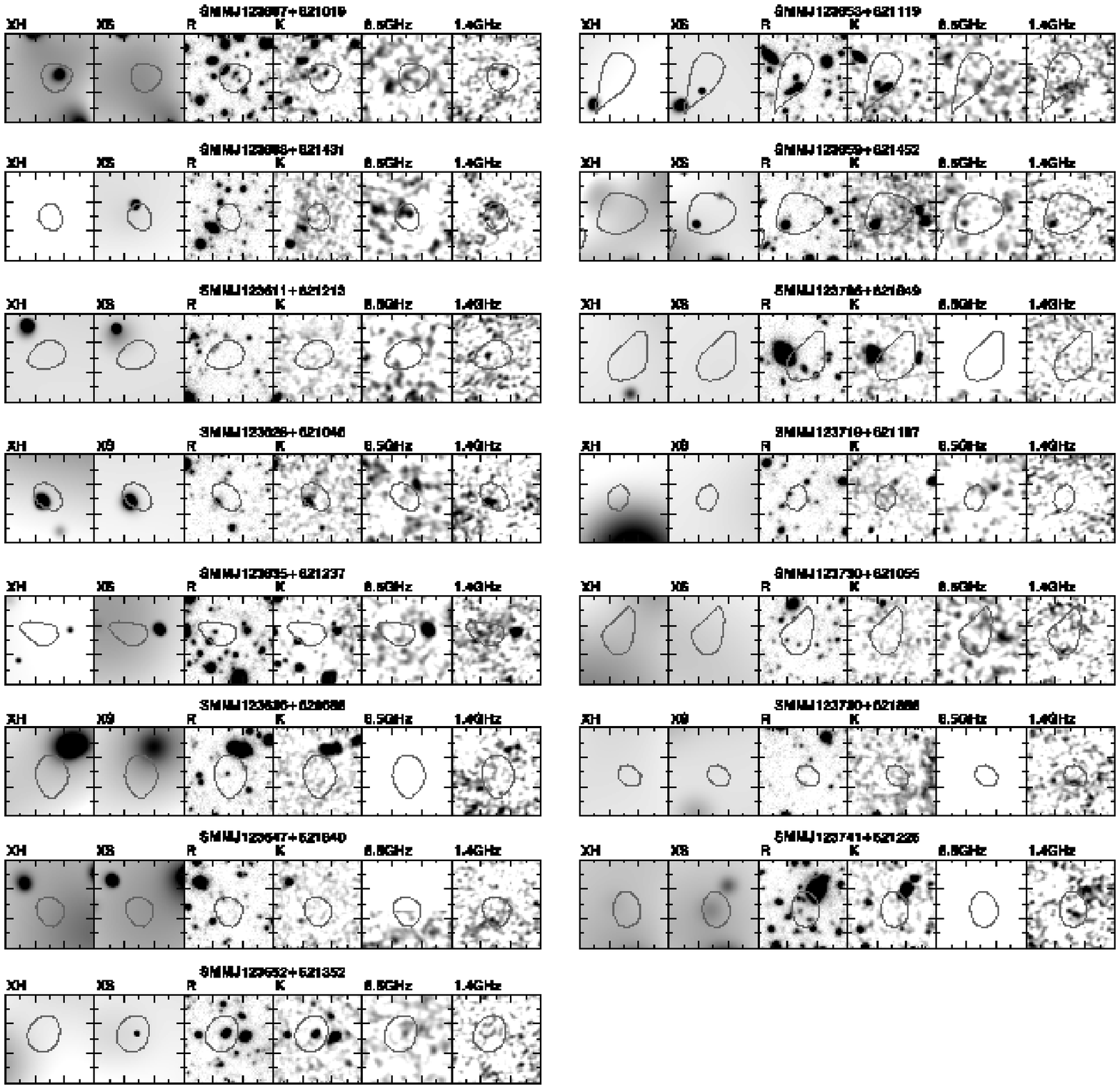}
\caption{Postage stamps of the $3.5\sigma <$SNR$<4.0\sigma$ 850\mum\ SCUBA
sources. These are the 17 sources from this supplementary catalogue. The images
are as described in Fig.~\ref{fig:lw4postage}. }
\label{fig:lw3postage}
\end{figure*}

In this sub-section we describe the 15 objects in the supplementary HDF-N SCUBA
catalogue of sources detected with a SNR between $3.5\sigma$ and
$4\sigma$.  The postage stamps are shown in Fig.~\ref{fig:lw3postage}.

\noindent\textbf{SMMJ123607+621019 [S]:}  A \chandra\ detected radio galaxy
seems the obvious ID, although the radio/sub-mm flux density
ratio is too low for the
redshift of 0.47 (determined from optical photometry).   The optical emission
suggests a disturbed system, and the surrounding area is very dense with
galaxies. Indeed, \citet{2002AJ....124.1839B} note that the photometric
redshift may be contaminated by flux from neighbouring galaxies.

\noindent\textbf{SMMJ123608+621431 [JS]:} The ID is likely to be the X-ray and
radio detected object to the North. 

\noindent\textbf{SMMJ123611+621213 [S]:} The radio map shows a $4\sigma$ source
4\arcs\ East of the SCUBA position, but with no optical counterpart.  
Although deeper observations may confirm this radio identification, for
now we conclude that there is no counterpart. 

\noindent\textbf{SMMJ123628+621046 [JS]:} The $z=1.013$ very red object that 
has detections in the optical, radio, and X-ray is the counterpart.

\noindent\textbf{SMMJ123635+621237 [JSP]:} This faint sub-mm source is
HDF\,850.7 in the  \citet{2003MNRAS.344..887S} catalogue.  It lies in a complex
field, with the most appealing candidate being a $z=1.219$ red galaxy.
This has detections in the radio, mid-IR, and X-ray.  Unfortunately this
candidate lies 7.6\arcs\ to the West of the SCUBA centre,
but again this may be due to confusion noise.  The redshift is compatible with
that derived from the CY estimator.  Despite the slightly large offset, we
will select this as the counterpart.

\noindent\textbf{SMMJ123636+620658 [S]:}  There is a very bright (presumably
foreground) galaxy with radio emission to the North.  A fainter source next to
it appears distorted (and perhaps redder).
From this fainter object there is radio
emission and hard X-ray flux.  However, this radio source is distant (over
11\arcs\ away), and sub-mm confusion is not a big issue here, since the source
is so bright.  We therefore conclude there is no counterpart.

\noindent\textbf{SMMJ123647+621840 [S]:}  This is also scan-map only (like
the previous source), but near the northern edge of the map.  It is apparently
a blank field to the limit of the observations. 

\noindent\textbf{SMMJ123652+621352 [JSP]:} 3 moderate redshift ISO objects are
within 11\arcs, but the SCUBA centre lies on top of non-ISO detected 
$z=1.355$ galaxy which also has \chandra\ and 8.5\,GHz
detections.  This is the only case where we base the radio ID on the 8.5\,GHz
data and not the 1.4\,GHz.  \citet{2003MNRAS.344..887S} note that this source
(HDF\,850.8) is part of an interacting pair of galaxies.

\noindent\textbf{SMMJ123653+621119 [SP]:} There are too many optically detected
sources present to choose from.  The nearby soft X-ray source does not have any
detectable radio flux, but is present in the X-ray study of star-forming 
galaxies by \citet{2002ApJ...576..625N}.  This Balmer-break galaxy has a 
redshift of 0.89 and an inferred SFR of 30$\,M_\odot\,{\rm yr}^{-1}$.
The small sub-mm flux density
is still moderately high for this SFR, and without a radio detection
we can only speculate that this is the correct ID.

\noindent\textbf{SMMJ123659+621452 [SP]:} 7\arcs\ to the South-east is a
$z=0.762$  X-ray, 1.4\,GHz radio and \textsl{ISO} detected
object.  7\arcs\ to the West lies a $z=0.849$ 8.5\,GHz radio and 
\textsl{ISO}
detected object. This latter object also has a faint 1.4\,GHz flux.  Both
objects have very similar radio and 15\mum\ fluxes, and hence both predict a
similar redshift when using the 850\mum\ fluxes as a photometric redshift
indicator. 

\noindent\textbf{SMMJ123706+621849 [S]:} It is difficult to choose among the 3
(at least) optical sources in the region, especially in the absence of any
other multi-wavelength data. Note the extended soft X-ray flux overlapping the
850\mum\ contours.

\noindent\textbf{SMMJ123719+621107 [JS]:} There are 4 optically detected
galaxies within 7\arcs\ of the SCUBA position, none of which have a radio
detection that can be used to help discern if any are the SCUBA counterpart.

\noindent\textbf{SMMJ123730+621055 [S]:} No obvious counterpart.

\noindent\textbf{SMMJ123731+621855 [S]:} No obvious counterpart.

\noindent\textbf{SMMJ123741+621225 [S]:} There is a radio source 7.6\arcs\ to
the North.  It overlaps with soft X-ray flux as seen in the \chandra\
image, although it is not formally listed in the Alexander et al.~(2003)
\chandra\ catalogue. The SCUBA contours prefer a very faint galaxy pair which
appears to have some radio flux.

\subsection{450$\bmath{\mu}$m detections $\bmath{>3.5\sigma}$}
\begin{figure*}
\includegraphics[width=7in,angle=0]{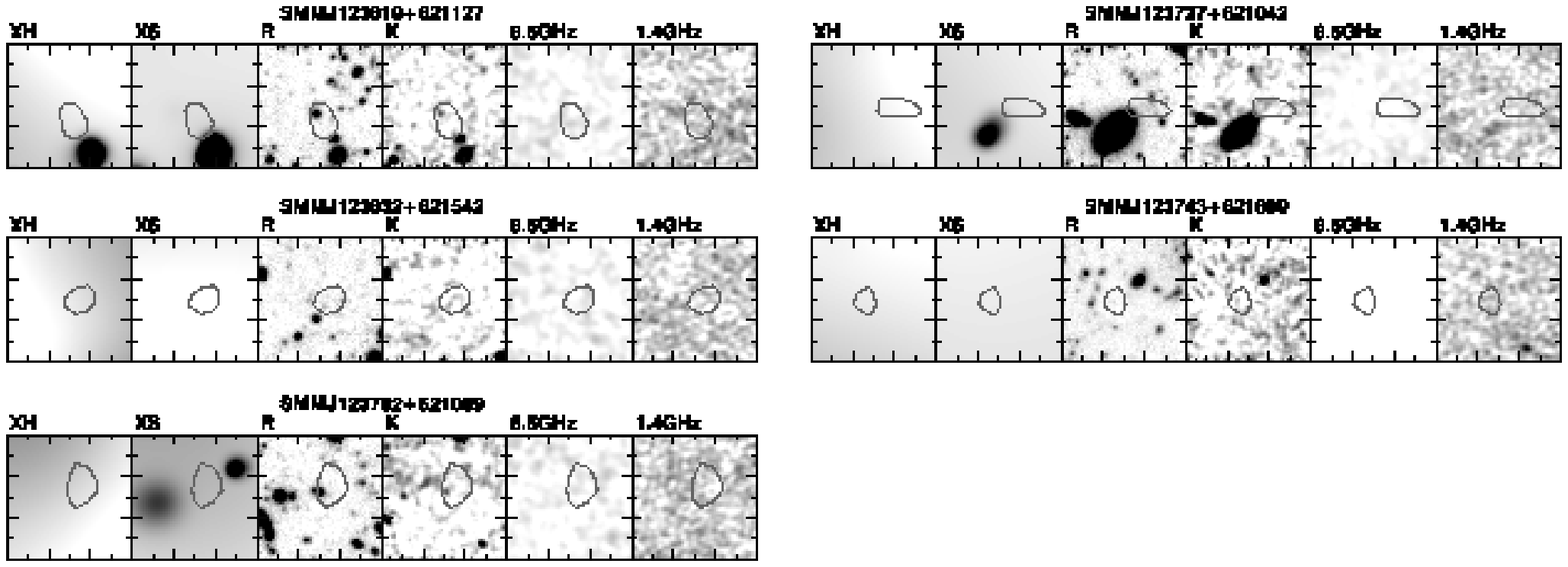}
\caption[Postage stamps of the $>4\sigma$ 450\mum\ SCUBA sources.]{Postage
stamps of the $>4\sigma$ 450\mum\ SCUBA objects.}
\label{fig:sw4postage}
\end{figure*}

We now describe the 5 objects in the HDF-N SCUBA map that were
detected at $>4\sigma$ confidence at 450\mum.  Here we search for counterparts
within 4\arcs\ of the SCUBA centroid.  Although the beam-size is smaller at
450\mum\ (7.5\arcs\,), the search radius still needs to be reasonably large,
because of the other systematic effects described in Section \ref{sec:statid}.
 It is also worth noting that anything detected
at 450\mum\ should have been seen at 850\mum, unless the source has an
extremely steep emissivity ($\beta>3.5$).  We are much less confident 
about the reliability of our 450\mum\ candidates, but nevertheless present
them for completeness (see also Fig.~\ref{fig:sw4postage}).

\noindent\textbf{SMMJ123619+621127 [JS]:} Since there is only 1 optically
detected galaxy within the search radius, we assign it as the tentative ID. 

\noindent\textbf{SMMJ123632+621542 [JS]:} No obvious counterpart.

\noindent\textbf{SMMJ123702+621009 [JS]:} 2 optical galaxies relatively near to
the centroid make it impossible to tell which is the counterpart. 

\noindent\textbf{SMMJ123727+621042 [S]:} A large elliptical galaxy is
nearby, but still too far to have it be the unambiguous ID.

\noindent\textbf{SMMJ123743+621609 [S]:}  No obvious counterpart.

\section{Other interpretations}
Using existing and new sub-mm data, \citet{rant} have recently presented 
another SCUBA map of the HDF-N.  They provided a catalog of 17 $>4\sigma$ 
objects\footnote{\cite{rant} label their objects GOODS850-01 through GOODS850-17, while we
use the prefix SMMJ combined with the coordinates of the object}, as well as
several more between $3-4\sigma$.  Given the strong interest in the GOODS-N region by many in the 
community, we briefly discuss the similarities and differences between the different analyses.

\subsection{Source catalogues}
Ignoring 3 Super-map sources that were found in regions sampled mainly in the scan-map mode, we find that 
14 of 16 $4\sigma$ Super-map sources are recovered in the list provided by \citet{rant}. 
SMMJ123701+621146 (a.k.a. HDF850.6) and SMMJ123702+621301 are not detected in their list because they use less
data and are hence less sensitive.  A full comparison is presented in Table~\ref{tab:rant}.

\begin{table}
\caption{Comparison between sources in Paper I and those reported in \citet{rant}. 
 Here we compare the fluxes and offsets between sources in our Super-map (B03),
and those in \citet{rant}.  Some scan-map sources are discrepant, but see text for an explanation.
}
\label{tab:rant}
\begin{tabular}{llllll}
\hline
\multicolumn{2}{c}{ID}& $\theta$ & \multicolumn{2}{c}{S850 (mJy)}  & \\
\multicolumn{1}{c}{B03} & \multicolumn{1}{c}{W04} & (\arcs\,) &  \multicolumn{1}{c}{B03} & \multicolumn{1}{c}{W04} &\\\hline
J123607+621143  & 31 &  18 & $15.2 \pm 3.9$ & $4.4  \pm 1.3$ & $^a$\\
J123608+621249  & 40 &  10 & $16.0 \pm 3.7$ & $3.9  \pm 1.3$ & $^a$\\
J123616+621516  & 07 &  4  & $6.1  \pm 0.9$ & $6.2  \pm 1.0$ & \\
J123618+621007  & 24 &  4  & $6.6  \pm 1.5$ & $6.0  \pm 1.7$ & \\
J123618+621552  & 03 &  1  & $7.2  \pm 0.9$ & $7.7  \pm 1.0$ & \\
J123621+621252  & 14 &  26 & $12.1 \pm 2.6$ & $10.5 \pm 2.3$ & $^a$\\
J123621+621710  & 15 &  2  & $8.8  \pm 1.5$ & $8.7  \pm 2.0$ & \\
J123622+621616  & 02 &  5  & $8.6  \pm 1.0$ & $10.3 \pm 1.2$ & \\
J123634+621407  & 05 &  6  & $11.2 \pm 1.6$ & $12.9 \pm 2.1$ & \\ 
J123637+621155  & 04 &  13 & $7.0  \pm 0.8$ & $8.6  \pm 1.3$ & \\ 
J123645+621449  & 11 &  1  & $8.5  \pm 1.3$ & $10.8 \pm 2.2$ & \\
J123650+621316  & 10 &  5  & $2.9  \pm 0.4$ & $2.6  \pm 0.5$ & \\  
J123652+621225  & 01 &  1  & $5.9  \pm 0.3$ & $5.1  \pm 0.5$ & \\  
J123656+621201  & 12 &  0  & $3.7  \pm 0.4$ & $3.3  \pm 0.7$ & \\   
J123700+620910  & 16 &  5  & $8.6  \pm 2.1$ & $12.4 \pm 2.9$ & \\
J123707+621410  & 09 &  7  & $9.9  \pm 2.5$ & $7.0 \pm 1.2$ & \\
J123713+621204  & 13 &  2  & $6.1  \pm 1.4$ & $7.0 \pm 1.5$ & \\\hline
\end{tabular}
\medskip
\\
$^a$\,Scan-map sources\\
\end{table}

Of their 17 $4\sigma$ objects, 14 are already detected in our $4\sigma$ catalogue.  Of the missing 
3, GOODS850-06 is in a region where the Super-map has low sensitivity (which they supplemented 
with deeper jiggle-maps), another (GOODS850-17) is detected in our catalogue at 
just under $4\sigma$, and the final object (GOODS850-08) is in a suspicious region that we address shortly.  
It is encouraging that separate groups can reproduce similar sources using 
different techniques.  We do note that the correlation between sources detected  under $4\sigma$ is much weaker,
but one would expect this given the poor reliability of low signal-to-noise sources.

\subsection{Jiggle-map `noise-spike'}
The only significant differences seem to be around SMMJ123607+621143 and SMMJ123608+62124, which were
first detected in the scan-map of \citet{2002MNRAS.330L..63B}.  \citet{rant} claim
scan-map data are suspect since they fail to recover them in new jiggle-map data.  Their source catalogue
contains three objects between $3.0\sigma-3.5\sigma$, two more between $3.5-4.0\sigma$, and one 
$>4\sigma$ within a SCUBA array size centered on these sources. This is an unusually large number for 
a typical `blank field' SCUBA observation\footnote{The number counts at this flux level suggest there 
should be at most two $>3\sigma$ sources in an area the size of the SCUBA array.}, and none of them have a 
plausible radio counterpart, nor are coincident with any of the scan-map sources.

\begin{figure*}
\includegraphics[width=7in,angle=0]{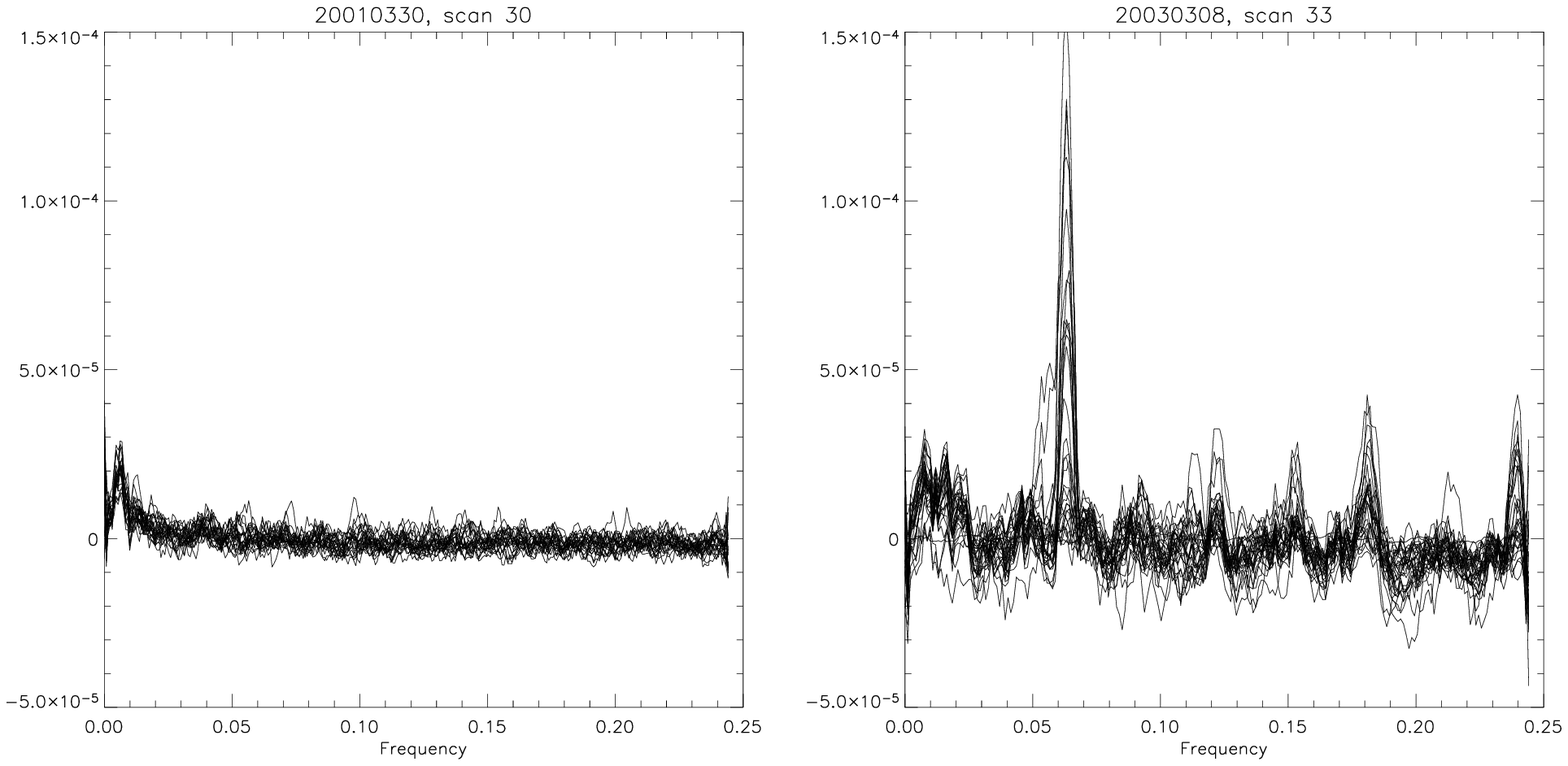}
\caption{The scan-synchronous noise glitch.  The left panel shows data taken while SCUBA was stable.
It is a collection of
power spectra taken from all bolometers, which is calculated using Fourier transforms of the raw data.  We label the abscissa `Frequency', although it is in fact the Fourier conjugate of the sample number.  Samples are collections of chopped data taken 16 seconds apart (this is how SCUBA records data). The right panel
clearly shows more insipid power spectra, especially at a period of 16 samples.  This corresponds to the same 
period on which the 16-point dither pattern operates, meaning noise can be projected onto the map at regular positions. }
\label{fig:glitch}
\end{figure*}

We believe some of the discrepancy arises due to the `noise-spike' issue discovered by one of us (Borys) in 
jiggle-map observations taken around the same time\footnote{Semesters 02B and 03A seem particularly
affected, the problem was noted in
other semesters to varying degrees as well.} (see Fig.~\ref{fig:glitch}).  This instrumental
fault has manifested itself in the map published in \citet{fls}, 
as well as in data from the SHADES survey (J. Dunlop, private communication).
Roughly $\sim2/3$ of the array is affected (24 bolometers) by noise that is scan-synchronous;  
i.e. the noise-spike occurs at the same frequency as the dither pattern cycle that makes up the jiggle-map.  Therefore the noise is projected into discrete places on the sky, and does not
integrate down as $t^{-1/2}$. Though it is known that the problem affects source fluxes and increases
the number false positives, there is currently no proven algorithm which removes this. \citet{rant} note 
their data are affected by this issue and removed a small number of especially noisy bolometers. Our own 
research has shown that this may not be sufficient, since during the sky subtraction phase of the data 
reduction, the bolometers affected by the noise corrupt the bolometers which are not. Fixes are being 
developed 
by our own group, the SHADES team (A. Mortier private communication), and T. Webb (private communication).  
IDL based code to identify the data affected by this issue is available from the author 
by request.

\begin{figure}
\includegraphics[width=3in,angle=0]{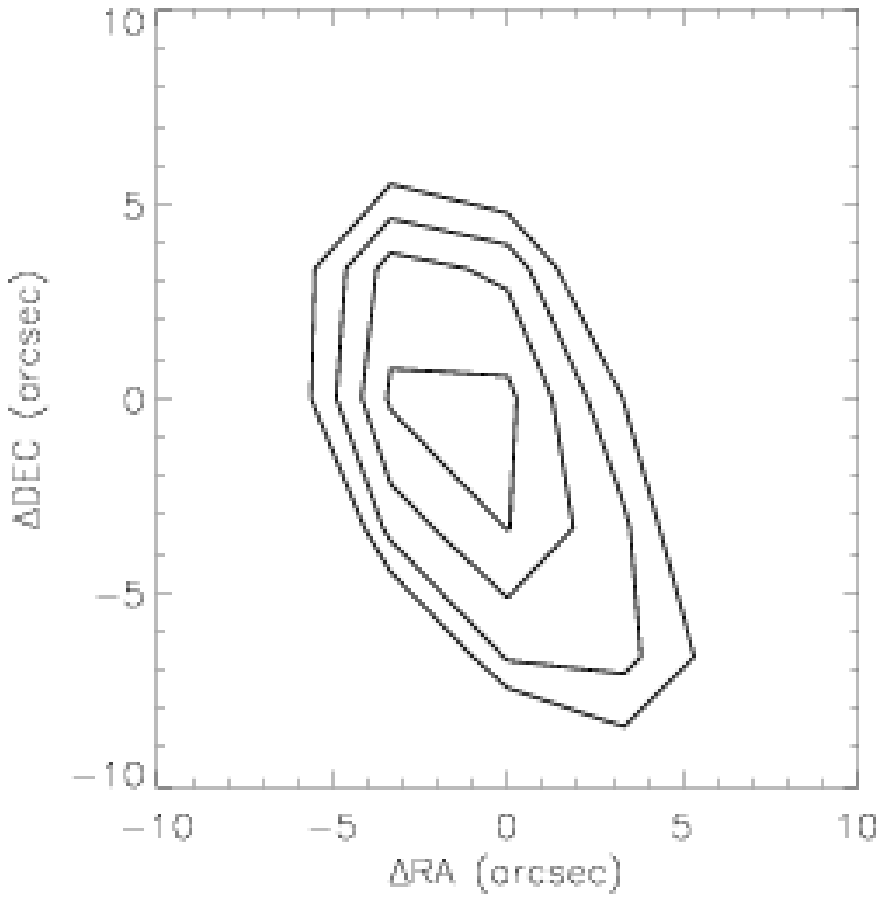}
\caption{Cross-correlation of maps made from scan- and jiggle- data separately. Contours are plotted at at $1,2,3,$ and $4\sigma$. These were calculated using pixels within 6\arcs\ of all objects in the
final $4\sigma$ catalogue, and demonstrate a very strong correlation between the two maps. }
\label{fig:scanjigg}
\end{figure}

Still, it useful to check to see how consistent scan-map and jiggle-map data are.  As described in Paper~I, we cross-correlated the jiggle and scan-maps in order to check the relative calibration
and pointing offsets (none of the `corrupt' jiggle-map data was used).  Fig.~\ref{fig:scanjigg} verifies that the two maps are sensitive to the same sources.  Given the success in using the scan-map mode for other projects \citep[e.g.][]{2000ApJS..131..505J}, there is no a priori reason to suspect that the scan-map technique is not viable. It is likely that a resolution between the recent results in \citet{rant} and our Super-map will require new observations from a well characterised and healthy sub-mm instrument.  Nevertheless, we stress that aside from this one region, the final source catalogues in \citet{rant}. and Paper~I are very consistent.

\subsection{Multi-wavelength identifications}
Regarding the multi-wavelength IDs, \citet{rant} find a similar radio association rate. 
Though by including HDF850.1 (SMMJ123652+621225) and
SMMJ123622+621616 (which is uncomfortably far away from the nearest radio source), they have 
estimated a higher radio-ID fraction than we found with our more rigorous identification recipe.
However, the most significant difference between the multi-wavelength IDs is
that \citet{rant} use X-ray or \textsl{ISO} detected objects that have no radio counterpart to tag
some SCUBA objects.  As we have demonstrated in this paper, unless there is a believable radio
source as the counterpart, we find no convincing evidence that an association can be made with another wavelength.

\end{document}